\documentclass[copyright,creativecommons]{eptcs}
\providecommand{\event}{GandALF 2014} 

\usepackage{cite}
\usepackage{array}
\usepackage[caption=false,font=footnotesize]{subfig}
\usepackage{dblfloatfix}
\usepackage{url}
\usepackage{tikz}
\usepackage{xcolor,pst-all}
\newgray{grayOne}{.9}
\newgray{gray3}{.8}

\newpsobject{showgrid}{psgrid}{subgriddiv=1,griddots=10,gridlabels=6pt}

\newcommand{\delete}[1]{}

\usepackage{epsfig}
\newcommand\text[1]{\mbox{#1}}
\usepackage{amssymb}

\title{Petri Games: \\ Synthesis of Distributed Systems with Causal Memory}
\author{Bernd Finkbeiner
\institute{Universit\"at des Saarlandes}
\email{finkbeiner@cs.uni-saarland.de}
\and
Ernst-R\"udiger Olderog
\institute{Carl von Ossietzky Universit\"at Oldenburg}
\email{olderog@informatik.uni-oldenburg.de}
}
\def\titlerunning{Petri Games}
\def\authorrunning{B. Finkbeiner and E.-R. Olderog}

\begin{document}

\newcommand\order{o}

\newcommand\comment[1]{}
\newtheorem{theorem}{Theorem}[section]
\newenvironment{Theorem}{\begin{theorem}\upshape}{\end{theorem}}
\newtheorem{corollary}[theorem]{Corollary}
\newenvironment{Corollary}{\begin{corollary}\upshape}{\end{corollary}}
\newtheorem{lemma}[theorem]{Lemma}
\newenvironment{Lemma}{\begin{lemma}\upshape}{\end{lemma}}
\newtheorem{proposition}[theorem]{Proposition}
\newenvironment{Proposition}{\begin{lemma}\upshape}{\end{lemma}}
\newtheorem{example}[theorem]{Example}
\newenvironment{Example}{\begin{lemma}\upshape}{\end{lemma}}
\newenvironment{proof}[1][Proof:]{\begin{trivlist}
\item[\hskip \labelsep {\bfseries #1}]}{\end{trivlist}}
\newcommand\qed{\hfill\ensuremath{\Box}}

\newcommand\fb{\lambda}

\newcommand\players{X}
\newcommand\environment{\mathit{env}}
\newcommand\placemarking{\mu}
\newcommand\bad{\mathcal{B}}

\newcommand\flow{\mathcal{F}}
\newcommand\places{\mathcal{P}}
\newcommand\transitions{\mathcal{T}}
\newcommand\initialmarking{\mathit{In}}
\newcommand\reach{\mathcal{R}}

\newcommand\Pset[1]{2^{#1}}

\newcommand\ident{\mathit{id}}
\renewcommand\wp{\mathit{wp}}
\newcommand\strategy{\mathit{strat}}
\newcommand\representation{\mathit{rep}}
\newcommand\forbidden{\mathit{forbidden}}

\newcommand\Mset{\mathbb{M}}
\newcommand\Nat{\mathbb{N}}

\newcommand\post[1]{\mathit{post}(#1)}
\newcommand\pre[1]{\mathit{pre}(#1)}

\newcommand\ms[2]{#1[#2]}  

\newcommand\pastnet[2]{#1|#2^-}
\newcommand\futurenet[2]{#1|#2^+}

\newcommand\maxslice[1]{#1 \mbox{}^\circ}
\newcommand\minslice[1]{\mbox{}^\circ #1}

\newcommand\transitionsplayer[1]{\rightarrow_{#1}}
\newcommand\fire[1]{[ #1\rangle}
\newcommand{\dotcup}{\ensuremath{\mathaccent\cdot\cup}}
\newcommand\enabled{\mathit{en}}
\newcommand\depends{\preceq}
\newcommand\observation{\mathit{obs}}

\newcommand\Venv{{V_{\mathit{env}}}}
\newcommand\dir{\mathit{dir}}
\newcommand\Alt{\mathit{Alt}}
\newcommand\child{\mathit{child}}
\newcommand\proj{\mathit{pr}}

\newcommand\ERO[1]{\marginpar{{\bf ERO:} #1}}
\newcommand\BF[1]{\marginpar{{\bf BF:} #1}}
\newcommand\past{\mathit{past}}
\newcommand\cut{\mathit{cut}}
\newcommand\reduce{\mathit{reduce}}
\newcommand\repeatstrategy{\mathit{repeat}}
\newcommand\down{\mathit{down}}
\newcommand\causalslice{\mathit{cut}}
\newcommand\causalrun{\mathit{cr}}

\newcommand\fu{\mathit{future}}

\newcommand\seqcomp{\,;}

\maketitle

\begin{abstract}

We present a new multiplayer game model for the interaction and the flow of information in a distributed system. The players are tokens on a Petri net. As long as the players move in independent parts of the net, they do not know of each other; when they synchronize at a joint transition, each player gets informed of the causal history of the other player.
We show that for Petri games with a single environment player and an arbitrary
bounded number of system players, deciding the existence of a safety strategy for the system players is EXPTIME-complete.

\end{abstract}

\section{Introduction}

Games are a natural model of the interaction between a computer system and its environment. Specifications are interpreted as winning conditions, implementations as strategies. An implementation is correct if the strategy is \emph{winning}, i.e., it ensures that the specification is met for all possible behaviors of the environment. Algorithms that determine the winner in the game between the system and its environment can be used to determine whether it is possible to implement a specification (the \emph{realizability} question) and, if the answer is yes, to automatically construct a correct implementation (the \emph{synthesis} problem).

 We present a new game model for the interaction and the flow of information in a distributed system. The players are tokens on a Petri net. In Petri nets, causality is represented by the flow of tokens through the net. It is therefore natural to designate tokens also as the carriers of information. As long as different players move in
concurrent places of the net, they do not know of each other.  Only
when they synchronize at a joint transition, each player gets informed
of the history of the other player, represented by all places and
transitions on which the joint transition causally depends.
The idea is that after such a joint transition, a strategy for a player can take the
history of all other players participating in the joint transition into account.
Think of a workflow where a document circulates in a large organization 
with many clerks and has to be signed by everyone, endorsing it or not.
Suppose a clerk wants to make the decision whether or not to endorse
it depending on who has endorsed it already.  As long as the clerk does not see
the document, he is undecided.  Only when he receives the document, he sees all
previous signatures and then makes his decision.

We call our extension of Petri nets \emph{Petri games}.  The players
are organized into two teams, the system players and the environment
players, where the system players wish to avoid a certain 
``bad'' place (i.e., they follow a safety objective), while the
environment players wish to reach just such a place. To partition
the tokens into the teams, we label each place as belonging to either
the system or the environment. A token belongs to a team whenever it is on a place that belongs to the team.

In the tradition of  Zielonka's automata~\cite{Zielonka/95/Asynchronous}, Petri games model distributed systems with \emph{causal memory}, i.e., distributed systems where the processes memorize their causal history and
communicate it to each other during each synchronization~\cite{Gastin+others/04/Distributed,MadhusudanTY/05/MSO,Genest+others/ICALP2013/Asynchronous}. Petri games thus abstract from the concrete content of a communication in that we assume that the processes always exchange the \emph{maximal} possible information, i.e., their entire causal history.  This is useful at a design stage before the details of the interface have been decided and one is more interested in restricting \emph{when} a communication can occur (e.g., when a device is connected to its base station, while a network connection is active, etc.) than \emph{what} may be communicated. The final interface is then determined by the information actually used by the winning strategies, which is typically only a small fraction of the causal history. Note that even though we assume the players to communicate everything they know, the flow of information in a Petri game is far from trivial. At any point, the players of the Petri game may have a different level of knowledge about the global state of the game, and the level of informedness changes dynamically as a result of the synchronizations chosen by the players.


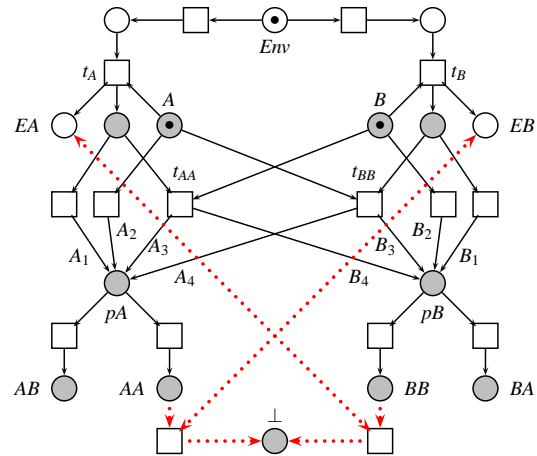
\begin{figure}[h]

\begin{minipage}{7cm}

Consider the development of a distributed security alarm system. If a burglar triggers the alarm at one location, the alarm should go off everywhere, and all locations should report the location where the original alarm occurred. This situation is depicted as a Petri net in Fig.~\ref{fig:intro-Petri-game}. The token that initially resides on place $\mathit{Env}$ represents the environment, which is, in our example, the burglar, who can decide to break into our building either at location A or B. The tokens that initially reside on places $A$ and $B$ represent the distributed controller consisting of two processes, the one on the left for location A and the one on the right for location B.
In the following, we will refer to the Petri net of Fig.~\ref{fig:intro-Petri-game} as a \emph{Petri game},  to emphasize that the tokens in fact represent players and that the nondeterminism present in the net is to be restricted by the (yet to be determined) strategy of the system players. 

\end{minipage}
\hspace{7mm}
\begin{minipage}{8cm}

\centering

 \psset{unit=1}

 \scalebox{0.7}{
\begin{pspicture}(0,0.5)(12,9)   
   \psset{arrows=->,nodesep=.25cm}

  \pscircle(6,9){.25} \pnode(6,9){p0}  \rput(6,9){$\bullet$}  \rput(6,8.5){$\mathit{Env}$}

  \pscircle(3,9){.25} \pnode(3,9){p1} 
  \pscircle(9,9){.25} \pnode(9,9){p2} 
  \psframe(4.25,8.75)(4.75,9.25) \pnode(4.5,9){t1}
  \psframe(7.25,8.75)(7.75,9.25) \pnode(7.5,9){t2} 
    
   \psframe(2.75,7.75)(3.25,8.25) \pnode(3,8){t21} \rput(2.5,8){$t_A$}
   \psframe(8.75,7.75)(9.25,8.25) \pnode(9,8){t22} \rput(9.5,8){$t_B$}
   
  \pscircle(2,7){.25} \pnode(2,7){p31}  \rput(1.3,7){$\mathit{EA}$}
  \pscircle[fillstyle=solid,fillcolor=lightgray](3,7){.25} \pnode(3,7){p32} 
  \pscircle[fillstyle=solid,fillcolor=lightgray](4,7){.25} \pnode(4,7){p33}  
                \rput(4,7){$\bullet$} \rput(4,7.5){$A$}
  \pscircle[fillstyle=solid,fillcolor=lightgray](8,7){.25} \pnode(8,7){p34}  
               \rput(8,7){$\bullet$} \rput(8,7.5){$B$}
  \pscircle[fillstyle=solid,fillcolor=lightgray](9,7){.25} \pnode(9,7){p35} 
  \pscircle(10,7){.25} \pnode(10,7){p36}  \rput(10.7,7){$\mathit{EB}$}

  \psframe(1.75,5.25)(2.25,5.75) \pnode(2,5.5){t41}
  \psframe(2.55,5.25)(3.05,5.75) \pnode(2.8,5.5){t42}
  \psframe(3.95,5.25)(4.45,5.75) \pnode(4.2,5.5){t43} \rput(4.3,6.1){$t_{AA}$}
  \psframe(7.55,5.25)(8.05,5.75) \pnode(7.8,5.5){t44} \rput(7.7,6.1){$t_{BB}$}
  \psframe(8.95,5.25)(9.45,5.75) \pnode(9.2,5.5){t45}
  \psframe(9.75,5.25)(10.25,5.75) \pnode(10,5.5){t46} 

  \pscircle[fillstyle=solid,fillcolor=lightgray](3,4){.25} \pnode(3,4){p51}
               \rput(3,3.4){$\mathit{pA}$}
  \pscircle[fillstyle=solid,fillcolor=lightgray](9,4){.25} \pnode(9,4){p52}  
               \rput(9,3.4){$\mathit{pB}$}

  \psframe(1.75,2.75)(2.25,3.25) \pnode(2,3){t61}
  \psframe(3.75,2.75)(4.25,3.25) \pnode(4,3){t62} 
  \psframe(7.75,2.75)(8.25,3.25) \pnode(8,3){t63}
  \psframe(9.75,2.75)(10.25,3.25) \pnode(10,3){t64} 
  
  \pscircle[fillstyle=solid,fillcolor=lightgray](2,2){.25} \pnode(2,2){p71}  
               \rput(1.3,2){$\mathit{AB}$}
  \pscircle[fillstyle=solid,fillcolor=lightgray](4,2){.25} \pnode(4,2){p72}
               \rput(3.3,2){$\mathit{AA}$}
  \pscircle[fillstyle=solid,fillcolor=lightgray](8,2){.25} \pnode(8,2){p73} 
               \rput(8.7,2){$\mathit{BB}$}
  \pscircle[fillstyle=solid,fillcolor=lightgray](10,2){.25} \pnode(10,2){p74}
               \rput(10.7,2){$\mathit{BA}$}


 \ncline[offset=0mm]{p0}{t1} \ncline[offset=0mm]{p0}{t2}
 \ncline[offset=0mm]{t1}{p1} \ncline[offset=0mm]{t2}{p2}
 \ncline[offset=0mm]{p1}{t21} \ncline[offset=0mm]{p2}{t22}

 \ncline[offset=0mm]{t21}{p31}
 \ncline[offset=0mm]{t21}{p32}
    \ncline[offset=0mm]{p33}{t21}

    \ncline[offset=0mm]{p34}{t22}
 \ncline[offset=0mm]{t22}{p35}
 \ncline[offset=0mm]{t22}{p36}

 \ncline[offset=0mm]{p32}{t41} \rput(2.3,4.5){$A_1$}
 \ncline[offset=0mm]{p32}{t43} \rput(3.2,5){$A_2$}
 \ncline[offset=0mm]{p33}{t42} \rput(3.8,4.7){$A_3$}
 \ncline[offset=0mm]{p33}{t44} \rput(4.3,4.1){$A_4$}

 \ncline[offset=0mm]{p34}{t43} \rput(7.6,4.1){$B_4$}
 \ncline[offset=0mm]{p34}{t45} \rput(8.1,4.7){$B_3$}
 \ncline[offset=0mm]{p35}{t44} \rput(8.8,5){$B_2$}
 \ncline[offset=0mm]{p35}{t46} \rput(9.7,4.5){$B_1$}

 \ncline[offset=0mm]{t41}{p51}
 \ncline[offset=0mm]{t42}{p51}
 \ncline[offset=0mm]{t43}{p51}
 \ncline[offset=0mm]{t44}{p51}

 \ncline[offset=0mm]{t43}{p52}
 \ncline[offset=0mm]{t44}{p52}
 \ncline[offset=0mm]{t45}{p52}
 \ncline[offset=0mm]{t46}{p52}

 \ncline[offset=0mm]{p51}{t61}
 \ncline[offset=0mm]{p51}{t62}
 \ncline[offset=0mm]{p52}{t63}
 \ncline[offset=0mm]{p52}{t64}

 \ncline[offset=0mm]{t61}{p71}
 \ncline[offset=0mm]{t62}{p72}
 \ncline[offset=0mm]{t63}{p73}
 \ncline[offset=0mm]{t64}{p74}


  \psframe(3.75,0.75)(4.25,1.25) \pnode(4,1){t81} 
  \psframe(7.75,0.75)(8.25,1.25) \pnode(8,1){t82}


  \pscircle[fillstyle=solid,fillcolor=lightgray](6,1){.25} \pnode(6,1){pbad}
               \rput(6,1.5){$\bot$}


 \ncline[offset=0mm,linestyle=dotted,arrows=<->,linewidth=2pt,linecolor=red]{p36}{t81} 
   \ncline[offset=0mm,linestyle=dotted,arrows=<->,linewidth=2pt,linecolor=red]{p31}{t82}
 \ncline[offset=0mm,linestyle=dotted,linewidth=2pt,linecolor=red]{p72}{t81} 
   \ncline[offset=0mm,linestyle=dotted,linewidth=2pt,linecolor=red]{p73}{t82}


\ncline[offset=0mm,linestyle=dotted,linewidth=2pt,linecolor=red]{t81}{pbad} 
   \ncline[offset=0mm,linestyle=dotted,linewidth=2pt,linecolor=red]{t82}{pbad}

\end{pspicture}
 } 

 
\caption{Introductory example of a Petri game 
modeling a distributed security alarm.
Places belonging to the system players $A$ and $B$ are shown in gray.
In the Petri game, the transitions to the bad place $\bot$ are shown with dotted lines. 
}
\label{fig:intro-Petri-game}

\end{minipage}

\end{figure}



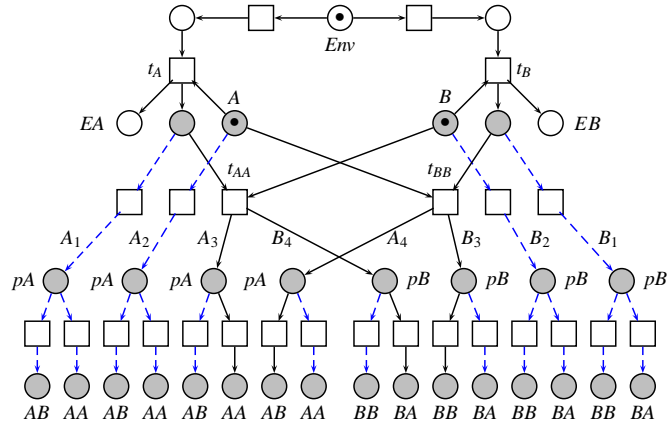
\begin{figure}[t]
\centering


 \psset{unit=1}

\scalebox{0.7}{
\begin{pspicture}(0,1)(12,9) 
   \psset{arrows=->,nodesep=.25cm}

 \pscircle(6,9){.25} \pnode(6,9){p0}  \rput(6,9){$\bullet$}  \rput(6,8.5){$\mathit{Env}$}

  \pscircle(3,9){.25} \pnode(3,9){p1} 
  \pscircle(9,9){.25} \pnode(9,9){p2} 
  \psframe(4.25,8.75)(4.75,9.25) \pnode(4.5,9){t1}
  \psframe(7.25,8.75)(7.75,9.25) \pnode(7.5,9){t2} 

  \psframe(2.75,7.75)(3.25,8.25) \pnode(3,8){t21} \rput(2.5,8){$t_A$}
   \psframe(8.75,7.75)(9.25,8.25) \pnode(9,8){t22} \rput(9.5,8){$t_B$}
   
  \pscircle(2,7){.25} \pnode(2,7){p31}  \rput(1.3,7){$\mathit{EA}$}
  \pscircle[fillstyle=solid,fillcolor=lightgray](3,7){.25} \pnode(3,7){p32} 
  \pscircle[fillstyle=solid,fillcolor=lightgray](4,7){.25} \pnode(4,7){p33}  
                \rput(4,7){$\bullet$} \rput(4,7.5){$A$}
  \pscircle[fillstyle=solid,fillcolor=lightgray](8,7){.25} \pnode(8,7){p34}  
               \rput(8,7){$\bullet$} \rput(8,7.5){$B$}
  \pscircle[fillstyle=solid,fillcolor=lightgray](9,7){.25} \pnode(9,7){p35} 
  \pscircle(10,7){.25} \pnode(10,7){p36}  \rput(10.7,7){$EB$}

  \psframe(1.75,5.25)(2.25,5.75) \pnode(2,5.5){t41}
  \psframe(2.75,5.25)(3.25,5.75) \pnode(3,5.5){t42}
  \psframe(3.75,5.25)(4.25,5.75) \pnode(4,5.5){t43} \rput(4.1,6.1){$t_{AA}$}
  \psframe(7.75,5.25)(8.25,5.75) \pnode(8,5.5){t44} \rput(7.9,6.1){$t_{BB}$}
  \psframe(8.75,5.25)(9.25,5.75) \pnode(9,5.5){t45}
  \psframe(9.75,5.25)(10.25,5.75) \pnode(10,5.5){t46} 


 \pscircle[fillstyle=solid,fillcolor=lightgray](0.6,4){.25} \pnode(0.6,4){p51}
                 \rput(0,4){$\mathit{pA}$}
 \pscircle[fillstyle=solid,fillcolor=lightgray](2.1,4){.25} \pnode(2.1,4){p52}
                   \rput(1.5,4){$\mathit{pA}$}
 \pscircle[fillstyle=solid,fillcolor=lightgray](3.6,4){.25} \pnode(3.6,4){p53}
                     \rput(3,4){$\mathit{pA}$}
 \pscircle[fillstyle=solid,fillcolor=lightgray](5.1,4){.25} \pnode(5.1,4){p54}
                      \rput(4.4,4){$\mathit{pA}$}

 \pscircle[fillstyle=solid,fillcolor=lightgray](6.85,4){.25} \pnode(6.85,4){p55}
                      \rput(7.5,4){$\mathit{pB}$}
 \pscircle[fillstyle=solid,fillcolor=lightgray](8.35,4){.25} \pnode(8.35,4){p56}
                    \rput(9,4){$\mathit{pB}$}
 \pscircle[fillstyle=solid,fillcolor=lightgray](9.85,4){.25} \pnode(9.85,4){p57}
                  \rput(10.5,4){$\mathit{pB}$}
 \pscircle[fillstyle=solid,fillcolor=lightgray](11.35,4){.25} \pnode(11.35,4){p58}
                      \rput(12,4){$\mathit{pB}$}


  \psframe(0,2.75)(0.5,3.25) \pnode(0.25,3){t511}
  \psframe(0.75,2.75)(1.25,3.25) \pnode(1,3){t512}

  \psframe(1.5,2.75)(2,3.25) \pnode(1.75,3){t521}
  \psframe(2.25,2.75)(2.75,3.25) \pnode(2.5,3){t522} 

  \psframe(3,2.75)(3.5,3.25) \pnode(3.25,3){t531}
  \psframe(3.75,2.75)(4.25,3.25) \pnode(4,3){t532}

  \psframe(4.5,2.75)(5,3.25) \pnode(4.75,3){t541}
  \psframe(5.25,2.75)(5.75,3.25) \pnode(5.5,3){t542}

 \pscircle[fillstyle=solid,fillcolor=lightgray](0.25,2){.25} \pnode(0.25,2){p511}
               \rput(0.25,1.5){$\mathit{AB}$}
 \pscircle[fillstyle=solid,fillcolor=lightgray](1,2){.25} \pnode(1,2){p512}
                \rput(1,1.5){$\mathit{AA}$}
 \pscircle[fillstyle=solid,fillcolor=lightgray](1.75,2){.25} \pnode(1.75,2){p521}
                 \rput(1.75,1.5){$\mathit{AB}$}
 \pscircle[fillstyle=solid,fillcolor=lightgray](2.5,2){.25} \pnode(2.5,2){p522}
                 \rput(2.5,1.5){$\mathit{AA}$}
 \pscircle[fillstyle=solid,fillcolor=lightgray](3.25,2){.25} \pnode(3.25,2){p531}
                \rput(3.25,1.5){$\mathit{AB}$}
 \pscircle[fillstyle=solid,fillcolor=lightgray](4,2){.25} \pnode(4,2){p532}
                \rput(4,1.5){$\mathit{AA}$}
 \pscircle[fillstyle=solid,fillcolor=lightgray](4.75,2){.25} \pnode(4.75,2){p541}
               \rput(4.75,1.5){$\mathit{AB}$}
 \pscircle[fillstyle=solid,fillcolor=lightgray](5.5,2){.25} \pnode(5.5,2){p542}
                \rput(5.5,1.5){$\mathit{AA}$}


  \psframe(6.25,2.75)(6.75,3.25) \pnode(6.5,3){t551}
  \psframe(7,2.75)(7.5,3.25) \pnode(7.25,3){t552}

  \psframe(7.75,2.75)(8.25,3.25) \pnode(8,3){t561}
  \psframe(8.5,2.75)(9,3.25) \pnode(8.75,3){t562} 

  \psframe(9.25,2.75)(9.75,3.25) \pnode(9.5,3){t571}
  \psframe(10,2.75)(10.5,3.25) \pnode(10.25,3){t572}

  \psframe(10.75,2.75)(11.25,3.25) \pnode(11,3){t581}
  \psframe(11.5,2.75)(12,3.25) \pnode(11.75,3){t582}

 \pscircle[fillstyle=solid,fillcolor=lightgray](6.5,2){.25} \pnode(6.5,2){p551}
               \rput(6.5,1.5){$\mathit{BB}$}
 \pscircle[fillstyle=solid,fillcolor=lightgray](7.25,2){.25} \pnode(7.25,2){p552}
                \rput(7.25,1.5){$\mathit{BA}$}
 \pscircle[fillstyle=solid,fillcolor=lightgray](8,2){.25} \pnode(8,2){p561}
                 \rput(8,1.5){$\mathit{BB}$}
 \pscircle[fillstyle=solid,fillcolor=lightgray](8.75,2){.25} \pnode(8.75,2){p562}
                 \rput(8.75,1.5){$\mathit{BA}$}
 \pscircle[fillstyle=solid,fillcolor=lightgray](9.5,2){.25} \pnode(9.5,2){p571}
                \rput(9.5,1.5){$\mathit{BB}$}
 \pscircle[fillstyle=solid,fillcolor=lightgray](10.25,2){.25} \pnode(10.25,2){p572}
                \rput(10.25,1.5){$\mathit{BA}$}
 \pscircle[fillstyle=solid,fillcolor=lightgray](11,2){.25} \pnode(11,2){p581}
               \rput(11,1.5){$\mathit{BB}$}
 \pscircle[fillstyle=solid,fillcolor=lightgray](11.75,2){.25} \pnode(11.75,2){p582}
                \rput(11.75,1.5){$\mathit{BA}$}


 \ncline[offset=0mm]{p0}{t1} \ncline[offset=0mm]{p0}{t2}
 \ncline[offset=0mm]{t1}{p1} \ncline[offset=0mm]{t2}{p2}
 \ncline[offset=0mm]{p1}{t21} \ncline[offset=0mm]{p2}{t22}

 \ncline[offset=0mm]{t21}{p31}
 \ncline[offset=0mm]{t21}{p32}
    \ncline[offset=0mm]{p33}{t21}

    \ncline[offset=0mm]{p34}{t22}
 \ncline[offset=0mm]{t22}{p35}
 \ncline[offset=0mm]{t22}{p36}

 \ncline[offset=0mm,linestyle=dashed,linecolor=blue]{p32}{t41}
 \ncline[offset=0mm]{p32}{t43} 
 \ncline[offset=0mm,linestyle=dashed,linecolor=blue]{p33}{t42} 
 \ncline[offset=0mm]{p33}{t44}

 \ncline[offset=0mm]{p34}{t43}
 \ncline[offset=0mm,linestyle=dashed,linecolor=blue]{p34}{t45} 
 \ncline[offset=0mm]{p35}{t44} 
 \ncline[offset=0mm,linestyle=dashed,linecolor=blue]{p35}{t46}

 \ncline[offset=0mm,linestyle=dashed,linecolor=blue]{t41}{p51}
         \rput(0.9,4.8){$A_1$}
 \ncline[offset=0mm,linestyle=dashed,linecolor=blue]{t42}{p52} 
         \rput(2.2,4.8){$A_2$}
 \ncline[offset=0mm]{t43}{p53} \rput(3.5,4.8){$A_3$}
 \ncline[offset=0mm]{t44}{p54} \rput(7.1,4.8){$A_4$}

 \ncline[offset=0mm]{t43}{p55} \rput(4.9,4.8){$B_4$}
 \ncline[offset=0mm]{t44}{p56}  \rput(8.5,4.8){$B_3$}
 \ncline[offset=0mm,linestyle=dashed,linecolor=blue]{t45}{p57}  
         \rput(9.8,4.8){$B_2$}
 \ncline[offset=0mm,linestyle=dashed,linecolor=blue]{t46}{p58}  
         \rput(11.1,4.8){$B_1$}

 \ncline[offset=0mm,linestyle=dashed,linecolor=blue]{p51}{t511}
 \ncline[offset=0mm,linestyle=dashed,linecolor=blue]{p51}{t512}
 \ncline[offset=0mm,linestyle=dashed,linecolor=blue]{p52}{t521}
 \ncline[offset=0mm,linestyle=dashed,linecolor=blue]{p52}{t522}

 \ncline[offset=0mm,linestyle=dashed,linecolor=blue]{p53}{t531}
 \ncline[offset=0mm]{p53}{t532}
 \ncline[offset=0mm]{p54}{t541}
 \ncline[offset=0mm,linestyle=dashed,linecolor=blue]{p54}{t542}

 \ncline[offset=0mm,linestyle=dashed,linecolor=blue]{t511}{p511}
 \ncline[offset=0mm,linestyle=dashed,linecolor=blue]{t512}{p512}
 \ncline[offset=0mm,linestyle=dashed,linecolor=blue]{t521}{p521}
 \ncline[offset=0mm,linestyle=dashed,linecolor=blue]{t522}{p522}

 \ncline[offset=0mm,linestyle=dashed,linecolor=blue]{t531}{p531}
 \ncline[offset=0mm]{t532}{p532}
 \ncline[offset=0mm]{t541}{p541}
 \ncline[offset=0mm,linestyle=dashed,linecolor=blue]{t542}{p542}

 \ncline[offset=0mm,linestyle=dashed,linecolor=blue]{p55}{t551}
 \ncline[offset=0mm]{p55}{t552}
 \ncline[offset=0mm]{p56}{t561}
 \ncline[offset=0mm,linestyle=dashed,linecolor=blue]{p56}{t562}

 \ncline[offset=0mm,linestyle=dashed,linecolor=blue]{p57}{t571}
 \ncline[offset=0mm,linestyle=dashed,linecolor=blue]{p57}{t572}
 \ncline[offset=0mm,linestyle=dashed,linecolor=blue]{p58}{t581}
 \ncline[offset=0mm,linestyle=dashed,linecolor=blue]{p58}{t582}

 \ncline[offset=0mm,linestyle=dashed,linecolor=blue]{t551}{p551}
 \ncline[offset=0mm]{t552}{p552}
 \ncline[offset=0mm]{t561}{p561}
 \ncline[offset=0mm,linestyle=dashed,linecolor=blue]{t562}{p562}

 \ncline[offset=0mm,linestyle=dashed,linecolor=blue]{t571}{p571}
 \ncline[offset=0mm,linestyle=dashed,linecolor=blue]{t572}{p572}
 \ncline[offset=0mm,linestyle=dashed,linecolor=blue]{t581}{p581}
 \ncline[offset=0mm,linestyle=dashed,linecolor=blue]{t582}{p582}

\end{pspicture}
} 

 
\caption{Unfolding of the Petri game in 
Fig.~\ref{fig:intro-Petri-game}. 
To aid visibility, the transitions leading to $\bot$
are omitted from the unfolding.
If the transitions shown with dashed lines are removed from the unfolding, the resulting net is a winning strategy for the system players.
}
\label{fig:unfold-intro-Petri-game}

\end{figure}


The system players and the environment players move on separate places in the net, the places belonging to the system players are shown in gray. 
In the example, our goal is to find a strategy for the system players that avoids a \emph{false alarm}, i.e., a marking where the environment token is still on $\mathit{Env}$ and at least one system token is on one of the places at the bottom, i.e., $\mathit{AA}$, $\mathit{AB}$, etc., 
and a \emph{false report}, i.e., a marking where the environment token is on place $\mathit{EA}$ and some system token is on $\mathit{AB}$ or $\mathit{BB}$ or a marking where the environment token is on $\mathit{EB}$ and some system token is on $\mathit{AA}$ or $\mathit{BA}$. 
To identify such undesirable markings we introduce a 
distinguished place $\bot$.
Fig.~\ref{fig:intro-Petri-game} shows (dashed) transitions 
towards $\bot$ firing at two instances of false reports, 
when tokens are on both $\mathit{EA}$ and $\mathit{BB}$ or 
on both $\mathit{EB}$ and $\mathit{AA}$. Similar transitions for other erroneous situations are omitted here to aid visibility.

Suppose that, in our Petri game, the burglar breaks into location $A$ by taking the left transition. 
Once the system token in $A$ has recorded this via transition $t_A$, it
has two possibilities: either synchronize with the system token in $B$ by taking transition $t_{\mathit{AA}}$, or skip the communication and go straight to $\mathit{pA}$ via transition $A_1$. Intuitively, only the choice to synchronize is a good move, because the system token in $B$ has no other way of hearing about the alarm. 
The only remaining move for the system token in $B$ would be to move ``spontaneously'' via transition $B_2$ to $pB$, at which point it would need to move to $\mathit{BA}$, because the combination of \emph{BB} and \emph{EA} would constitute a false alarm. However, the token in $pB$ has no way of distinguishing this situation from one where the environment token is still on $\mathit{Env}$; in this situation, the move to $\mathit{EA}$ would also reach a false alarm.

Our definition of strategies is based on the \emph{unfolding} of the net, which is shown for our example in Fig.~\ref{fig:unfold-intro-Petri-game}. By eliminating all joins in the net, \emph{net unfoldings} \cite{NPW81,Eng91,EH08} separate places that are reached via multiple causal histories into separate copies.
In the example, place $\mathit{pB}$ has been unfolded into four separate copies, corresponding to the four different ways to reach $\mathit{pB}$, via the transition arcs $B_1$ through $B_4$. Each copy represents different knowledge:
in $B_1$, only $B$ knows that there has been a burglary at location $B$;
in $B_2$, $B$ knows nothing;
in $B_3$, $B$ knows that $A$ knows that there has been a burglary at position $B$;
in $B_4$, $B$ knows that there has been a burglary at location $A$.
(Symmetric statements hold for $pA$ and the transition arcs $A_1$ -- $A_4$.)
In the unfolding, it becomes clear that taking transition $B_2$ is a bad move, because reaching the bad marking containing $\mathit{Env}$ and either $\mathit{BA}$ or $\mathit{BB}$ has now become unavoidable. A \emph{strategy} is a subprocess of the unfolding that preserves the local nondeterminism of the environment token. Fig.~\ref{fig:unfold-intro-Petri-game} shows a winning strategy for the system players: by omitting the dashed arrows, they can make the bad place $\bot$ unreachable and therefore win the game.

We show that for a single environment token and an arbitrary (but bounded) number of system tokens, deciding the existence of a safety strategy for the system players is EXPTIME-complete. This means that as long as there is a single source of information, such as the \emph{input} of an algorithm or the \emph{sender} in a communication protocol, solving Petri games is no more difficult than
solving standard combinatorial games under complete information~\cite{DBLP:journals/siamcomp/StockmeyerC79}.
The case of Petri games with two or more environment tokens, i.e., situations with two or more \emph{independent} information sources, remains open.

The remainder of the paper is structured as follows. 
In Section~\ref{section:games} we introduce the notion
of Petri games and define strategies based on net unfoldings. 
In Section~\ref{section:distribution} we show that for concurrency
preserving games every strategy can be distributed over local
controllers.
In Section~\ref{section:finiteness} we introduce 
the new  notion of mcuts on net unfoldings.
In Section~\ref{section:algorithm} we show that
the problem of deciding the winner of a Petri game is EXPTIME-complete.
Related work and conclusions are presented
in Sections~\ref{section:related} and \ref{section:conclusion}.
Due to space limitations, proofs have been moved into the full version of this paper.

\section{Petri nets}
 \label{section:nets}

We recall concepts from Petri net theory
\cite{Rei85,BF88,NPW81,Eng91,Esp94,MMS96,KKV03,EH08}.
A \emph{place/tran\-si\-tion} (\emph{P/T}) \emph{Petri net} 
or simply \emph{net} 
$\mathcal N = (\places, \transitions, \flow, \initialmarking)$ 
consists of 
possibly infinite, disjoint sets 
$\places$ of \emph{places} and $\transitions$ of transitions, 
a \emph{flow relation} $\flow$, which is a multiset over 
$(\places \times \transitions) \cup (\transitions \times \places)$, 
and an \emph{initial marking} $\initialmarking$.
In general, a \emph{marking} of $\mathcal N$ is a finite multiset over $\places$. 
It represents a global state of $\mathcal N$.
By convention, a net 
named $\mathcal N$ has
the components $\mathcal N = (\places, \transitions, \flow, \initialmarking)$,
and analogously for nets with decorated names like
$\mathcal{N}_1, \mathcal{N}_2, \mathcal{N}^U$.

The elements of $\places \cup \transitions$ are called 
\emph{nodes} of $\mathcal N$, thereby referring
to the  bipartite graphic representation of nets, 
where places are drawn as circles and transitions as boxes.
The flow relation $\flow$ is represented by directed arrows between
places and transitions. An arrow from a place $p$ to a 
transition $t$ is decorated by a
\emph{multiplicity} $k$ if $\flow(p,t) = k$, and analogously,
an arrow from a transition $t$ to a place $p$ 
is decorated by a
\emph{multiplicity} $k$ if $\flow(t,p) = k$.
We use a double arrow arc between a place and a transition
if there are arcs in both directions.
A marking $M$ is represented by placing $M(p)$ \emph{tokens}
in every place $p$.

$\mathcal N$ is \emph{finite} if it has only finitely many nodes,
and \emph{infinite} otherwise.
For nodes $x, y$ we write $x\, \flow\, y$ if $\flow(x,y) > 0$. 
The \emph{precondition} of $y$ is the multiset $\pre{y}$ over nodes
defined by $\pre{y}(x)=\flow(x,y)$.
The \emph{postcondition} of $x$ is the multiset $\post{x}$ over nodes
defined by  $\post{x}(y)=\flow(x,y)$.
When stressing the dependency on the net $\mathcal N$,
we write $pre^{\mathcal N}(y)$ and $post^{\mathcal N}(x)$
instead of $\pre{y}$ and  $\post{x}$.
As in \cite{Eng91} we require \emph{finite synchronization} \cite{BF88} 
and non-empty pre- and postconditions:
$\pre{t}$ and $\post{t}$ are finite, non-empty multisets
for all transitions $t \in \transitions$. 

A transition $t$ is \emph{enabled} at a marking $M$
if the multiset inclusion $\pre{t} \subseteq M$ holds.
\emph{Executing} or \emph{firing} such a transition $t$ at $M$ 
yields the successor marking $M'$ defined by 
$M' = M - \pre{t} + \post{t}$. We denote this by $M \fire{t} M'$.
The set of \emph{reachable markings} of a net $\mathcal N$
is denoted by $\reach(\mathcal{N})$ and defined by
$\reach(\mathcal{N}) = 
\{M \mid \exists\ t_1, \ldots, t_n \in\ \transitions:\ 
\initialmarking \fire{t_1}M_1 \fire{t_2} \ldots \fire{t_n} M_n = M \}$. 
A net $\mathcal N$ is $k$-\emph{bounded} for a given $k \in \mathbb{N}$
if $M(p) \leq k$ holds for all $M \in \reach(\mathcal{N})$ and all $p \in \places$.
It is \emph{bounded} if it is $k$-bounded for some given $k$
and  \emph{safe} if it is 1-bounded.

$\flow^+$ denotes the transitive closure  
and  $\flow^*$ the reflexive, transitive closure of $\flow$.
Nodes $x$ and $y$  are \emph{in conflict}, abbreviated by $x \sharp y$, 
if there exists a place $p \in \places$, different from $x$ and $y$,
from which one can reach $x$ and
$y$ via $\flow^+$, exiting $p$ by different arcs.
A node $x$ is in \emph{self-conflict} if $x \sharp x$.

We use the notations $\minslice{\mathcal N} = \{ p \in \places \mid \pre{p}=\emptyset\}$ 
and 
$\maxslice{\mathcal N}=\{ p \in \places  \mid \post{p}=\emptyset\}$
for the sets of places without incoming or
outgoing transitions, respectively.
For a multiset $M$ over $\places$ let
$\mathcal N[M]$ result from $\mathcal N$ by changing 
its initial marking $\initialmarking$ to $M$.
For a set $X$ of nodes we define the 
\emph{restriction} of $\mathcal N$ to $X$
as the net 
$\mathcal N \upharpoonright X = 
(\places \cap X, \transitions \cap X, 
\flow \upharpoonright (X \times X), 
\initialmarking \upharpoonright X)$.

Consider two nets $\mathcal N_1$ and $\mathcal N_2$.
Then $\mathcal N_1$ is an \emph{initial subnet} or simply
\emph{subnet} of $\mathcal N_2$,
denoted by $\mathcal N_1 \sqsubseteq \mathcal N_2$, if
$\places_1 \subseteq \places_2$,
$\transitions_1  \subseteq \transitions_2$,
$\flow_1 \subseteq \flow_2$, and
$\initialmarking_1 = \initialmarking_2$.
A \emph{homomorphism} from $\mathcal N_1$ to $\mathcal N_2$ is a mapping 
$h: \places_1 \cup \transitions_1 \rightarrow \places_2 \cup \transitions_2$ with
$h(\places_1) \subseteq \places_2$ and
$h(\transitions_1) \subseteq \transitions_2$,
and with 
$\forall t \in \transitions_1:$
$\ms{h}{\pre{t}} = \pre{h(t)}$ and 
 $\ms{h}{\post{t}} = \post{h(t)}$.
If additionally $\ms{h}{\initialmarking_1} = \initialmarking_2$, 
then $h$ is called an \emph{initial} homomorphism.
An (\emph{initial}) \emph{isomorphism} is a bijective (initial) homomorphism.

\smallskip

\textbf{Occurrence nets and unfoldings.}
To represent the occurrences of transitions
with both their causal dependency and 
conflicts (nondeterministic choices),
we consider occurrence nets, branching processes, and
unfoldings of Petri nets as in \cite{NPW81,Eng91,KKV03,EH08}.
We follow the axiomatic presentation in \cite{Eng91}, 
taking \cite{MMS96} into account for dealing with P/T Petri nets.

An \emph{occurrence net} is a Petri net
$\mathcal N$, where 
$\forall\, t \in \transitions: \pre{t}$ and $\post{t}$ are sets,
$\forall\, p \in \places: |\pre{p}| \le 1$,
the inverse flow relation $\flow^{-1}$ is well-founded,
no transition $t \in \transitions$ is in self-conflict,
and $\initialmarking = \minslice{\mathcal N}$.
Note that an occurrence net is a safe net.
Two nodes $x, y$ of an occurrence net are \emph{causally related} 
if $x\, \flow^*\, y$  or $y\, \flow^*\, x$. 
They are \emph{concurrent} if they are neither causally related 
nor in conflict.
If $x\, \flow^+\, y$ then $x$ is called a \emph{causal predecessor} of~$y$, abbreviated $x<y$. We write $x\le y$ if $x<y$ or $x=y$.
The \emph{causal past} of a node $y$ is the set 
$\past(y)=\{x \mid x \le y\}$.

A \emph{branching process} of a net $\mathcal N$ 
is a pair $\beta = (\mathcal N^U, \fb)$, where $\mathcal N^U$ 
is an occurrence net and $\fb$ is a ``labeling'', i.e., a homomorphism 
from $\mathcal N^U$ to $\mathcal N$ that
is injective on transitions with the same precondition:
$\forall\, t_1, t_2 \in \transitions^U:
       \pre{t_1} = \pre{t_2} \land \fb(t_1) = \fb(t_2)$
       implies $t_1 = t_2$.
If $\fb$ is initial, $\beta$ is called
an \emph{initial branching process}.
The \emph{unfolding} of a net $\mathcal N$ 
is an initial branching process $\beta_U =(\mathcal N^U, \fb)$ that is
\emph{complete} in the sense that every transition of the net 
is recorded in the unfolding:
      $ \forall\, t \in \transitions, \forall\, C \subseteq \places^U$:
      if $C$ is a set of concurrent places and 
      $\ms{\fb}{C} = \pre{t}$,
      then there exists a transition $t^U \in  \transitions^U$ 
      such that $\pre{t^U} = C$ and $\fb(t^U) = t$.

Let $\beta_1 = (\mathcal N_1, \fb_1)$ and 
$\beta_2 = (\mathcal N_2, \fb_2)$ be two branching processes of
$\mathcal N$. A homomorphism from $\beta_1$ to $\beta_2$ is a
homomorphism $h$ from $\mathcal N_1$ to $\mathcal N_2$ 
with
$\fb_1 = \fb_2 \circ h$. 
It is called \emph{initial}
if $h$ is initial; it is an \emph{isomorphism} if $h$ is an isomorphism.
$\beta_1$ and $\beta_2$ are \emph{isomorphic} if there exists an 
initial isomorphism from  $\beta_1$ to $\beta_2$.
$\beta_1$ \emph{approximates} $\beta_2$ if there exists an
initial injective homomorphism from  $\beta_1$ to $\beta_2$.
$\beta_1$ is a \emph{subprocess} of $\beta_2$ if
$\beta_1$ \emph{approximates} $\beta_2$ with the identity 
on $\places_1 \cup \transitions_1$ as the homomorphism. 
Thus $\mathcal N_1 \sqsubseteq \mathcal N_2$ and
$\fb_1 = \fb_2 \upharpoonright (\places_1 \cup \transitions_1)$.
If $\beta_1$ \emph{approximates} $\beta_2$ then 
$\beta_1$ is isomorphic to a subprocess of $\beta_2$.

In \cite{Eng91} is shown that the unfolding 
$\beta_U = (\mathcal N^U, \fb)$ of a net $\mathcal N$
is unique up to isomorphism and 
that every initial branching process $\beta_1$
of $\mathcal N$ approximates $\beta_U$.
Thus up to isomorphism we can assume that $\beta_1$
is a subprocess of $\beta_U$.

\smallskip

\textbf{Cuts and sequential composition.}
A \emph{cut} of an occurrence net $\mathcal N$ 
is a maximal subset of the places that are pairwise concurrent.
For a cut $C$ let 
$C^- = {\{x \in \places \cup \transitions \mid \exists s \in C:\ x \leq s \}}$
and 
$C^+ = \{x \in \places \cup \transitions \mid$ $\exists s \in C:\ s \leq x \}$.
A cut $C$ splits $\mathcal N$ into the two nets 
$\mathcal N \upharpoonright C^-$ and 
$(\mathcal N \upharpoonright C^+)[C]$;
it also splits a branching process $(\mathcal N, \fb)$
into two branching processes $(\mathcal N_1, \fb_1)$ and
$(\mathcal N_2, \fb_2)$, where
$\mathcal N_1 = \mathcal N \upharpoonright C^-$ and
$\mathcal N_2 = (\mathcal N \upharpoonright C^+)[C]$ and
$\fb_1 = \fb \upharpoonright C^-$ and
$\fb_2 = \fb \upharpoonright C^+$.

Two branching processes 
$(\mathcal N_1, \fb_1)$ and $(\mathcal N_2, \fb_2)$ 
of a given P/T Petri net are \emph{compatible} if
$\ms{\fb_1}{\maxslice{\mathcal N_1}} =
\ms{\fb_2}{\minslice{\mathcal N_2}}$.
Given two compatible branching processes
$(\mathcal N_1, \fb_1)$ and $(\mathcal N_2, \fb_2)$,
we can up to isomorphisms of $\mathcal N_1$ and of $\mathcal N_2$ 
assume that 
$\maxslice{\mathcal N_1} = \minslice{\mathcal N_2}$
and construct a unique branching process $(\mathcal N, \fb)$ with 
$\mathcal N \upharpoonright C^- = \mathcal N_1$ 
and 
$(\mathcal N \upharpoonright C^+)[C] = \mathcal N_2$,
and
$\fb \upharpoonright  C^- = \fb_1$ 
and 
$\fb \upharpoonright  C^+ = \fb_2$,
for the cut $C = \maxslice{\mathcal N_1} = \minslice{\mathcal N_2}$.
This branching process is the \emph{sequential composition} of 
$(\mathcal N_1, \fb_1)$ and $(\mathcal N_2, \fb_2)$, denoted by 
$(\mathcal N, \fb) = (\mathcal N_1, \fb_1) \seqcomp (\mathcal N_2 , \fb_2)$. 
If $(\mathcal N_1, \fb_1)$ is an initial branching process, 
then so is $(\mathcal N, \fb)$.

\smallskip

\textbf{Causal nets and concurrent runs.}
Executions of Petri nets are represented by causal nets
and concurrent runs as in \cite{NPW81,BF88}.
A \emph{causal net} is an occurrence net $\mathcal N$,
where  $\forall\, p \in \places: |\post{p}| \le 1$.
Thus in a causal net there are no (self-) conflicts.
A (\emph{concurrent}) \emph{run} or \emph{process} of $\mathcal N$ is 
a special case of a branching process $\beta_R = (\mathcal N^R, \rho)$, 
where $\mathcal N^R$ is a causal net.
If $\rho$ is initial,  $\beta_R$ is called an \emph{initial run}. 
Note that every initial run of $\mathcal N$ approximates the
unfolding $\beta_U = (\mathcal N^U, \fb)$ of $\mathcal N$.
Thus up to isomorphism we can assume the an initial run of $\mathcal N$
is a subprocess of $\beta_U$. 

The marking \emph{reached by} a finite initial run 
$\beta_R = (\mathcal N^R, \rho)$ of $\mathcal N$ 
is denoted by $\fire{\beta_R}$ and defined as the multiset 
$\fire{\beta_R} = \ms{\rho}{\maxslice{(\mathcal N^R)}}$. 
We remark that the set $\reach(\mathcal{N})$
of reachable markings of $\mathcal N$
can be obtained via the runs as follows:
$\reach(\mathcal{N}) = 
   \{\, \fire{\beta_R} \mid 
    \beta_R \mbox{ is a finite initial run of }
   \mathcal N\, \}$.

\section{Petri Games}
\label{section:games}

We wish to model games where the players
proceed independently of each other, without information of each others state, unless they explicitly communicate. To this end, we introduce Petri games,
defined as  place/transition (P/T) Petri nets, 
where the set of places is partitioned into a subset 
$\places_S$ belonging to the \emph{system players} and a subset $\places_E$ belonging to the \emph{environment}.
Additionally, the Petri game identifies a set $\bad$ of \emph{bad} places 
(from the point of view of the system), 
which indicate a victory for the environment. 
Formally, a Petri game is a structure
$\mathcal G = (\places_S, \places_E, 
\transitions, \flow, \initialmarking, \bad)$,
where the (\emph{underlying}) \emph{Petri net of the game} $\mathcal G$ is
$\mathcal N = (\places, \transitions, \flow, \initialmarking)$ with places $\places=\places_S \cup \places_E$.
Players are modeled by the tokens of $\mathcal N$.
Throughout this paper we stipulate that 
there is only one environment player.

\vspace{2mm}

\begin{figure}[h]

\begin{minipage}{9cm}
\begin{example}
\label{exam:same-dec-with-test}
Fig.~\ref{fig:same-dec-with-test} shows the underlying 
P/T net $\mathcal N$ of a small Petri game
for two system players in place $\mathit{Sys}$ and one environment player in place $\mathit{Env}$.
Environment places are white and system places are gray.
The environment chooses $A$ or $B$ by executing
one of the transitions $t_1$ or $t_2$.
The goal of the system players is to achieve the same decisions as $\mathit{Env}$,
i.e., both system players should choose $A'$ if $\mathit{Env}$ chooses $A$,
and $B'$ if $\mathit{Env}$ chooses $B$.
Without communication, the system players do not know which decision
the environment has taken. 
However, when both system players and 
the environment communicate by synchronizing via the
transitions $\mathit{test}_1$ or $\mathit{test}_2$, the system players learn
about the decision taken by the environment and can mimic it.
If $\mathit{test}_1$ was successful, they choose $A'$ via
transition $t_1'$, and 
if $\mathit{test}_2$ was successful, they choose $B'$ via transition
$t_2'$.
\qed
\end{example}
\end{minipage}
\hspace{7mm}
\begin{minipage}{6cm}


\begin{center}

 \psset{unit=1}

\scalebox{0.7}{
\begin{pspicture}(0,5)(8,9)   
   \psset{arrows=->,nodesep=.25cm}


\pscircle(4,9){.25} \pnode(4,9){p49} 
    \rput(4,9.5){$\mathit{Env}$} 
    \rput(4,9){$\bullet$}


 \psframe(0.75,7.75)(1.25,8.25) \pnode(1,8){t18} 
     \rput(0.5,8){$t_1$}

 \psframe(6.75,7.75)(7.25,8.25) \pnode(7,8){t78} 
     \rput(7.5,8){$t_2$}

  
\pscircle[fillstyle=solid](1,7){.25} 
   \pnode(1,7){p17} \rput(0.5,7){$A$}

 \psframe(2.25,6.75)(2.75,7.25) \pnode(2.5,7){t257} 
     \rput(2.5,7.5){$\mathit{test}_1$}

\pscircle[fillstyle=solid,fillcolor=lightgray](4,7){.25} 
   \pnode(4,7){p47} \rput(4,7.5){$\mathit{Sys}$}
   \rput(3.9,7){$\bullet$} \rput(4.1,7){$\bullet$}

 \psframe(5.25,6.75)(5.75,7.25) \pnode(5.5,7){t557} 
     \rput(5.5,7.5){$\mathit{test}_2$}

 \pscircle[fillstyle=solid](7,7){.25} 
   \pnode(7,7){p77} \rput(7.5,7){$B$}


 \pscircle(1,6){.25} \pnode(1,6){p16} \rput(0.4,6){$\mathit{EA}$}

 \psframe(2.75,5.75)(3.25,6.25) \pnode(3,6){t36} 
     \rput(2.5,6){$t'_1$}

 \psframe(4.75,5.75)(5.25,6.25) \pnode(5,6){t56} 
     \rput(5.5,6){$t'_2$}

 \pscircle(7,6){.25} \pnode(7,6){p76} \rput(7.6,6){$\mathit{EB}$}


\pscircle[fillstyle=solid,fillcolor=lightgray](3,5){.25} 
   \pnode(3,5){p35} \rput(3,4.5){$A'$}

 \pscircle[fillstyle=solid,fillcolor=lightgray](5,5){.25} 
   \pnode(5,5){p55} \rput(5,4.5){$B'$}


 \ncline[offset=0mm]{p49}{t18}  \ncline[offset=0mm]{p49}{t78}

 \ncline[offset=0mm]{t18}{p17} 
 \ncline[offset=0mm]{t78}{p77}

 \ncline[offset=0mm]{p17}{t257}
\ncline[offset=0mm]{t257}{p16}
\ncline[offset=0mm,arrows=<->]{p47}{t257}  \rput(3.2,7.2){2}

 \ncline[offset=0mm,arrows=<->]{p47}{t557}  \rput(4.8,7.2){2}
\ncline[offset=0mm]{p77}{t557}
\ncline[offset=0mm]{t557}{p76}

 \ncline[offset=0mm]{p47}{t36} \ncline[offset=0mm]{p47}{t56}

 \ncline[offset=0mm]{t36}{p35} 
 \ncline[offset=0mm]{t56}{p55}

\end{pspicture}
} 

\end{center}


\caption{Petri game for achieving same decisions,
where $\mathit{Env}$ and $\mathit{Sys}$ can synchronize via two transitions
$\mathit{test}_1$ and $\mathit{test}_2$. Transitions from $\mathit{EA}$ and $B'$ and from
$\mathit{EB}$ and $A'$ to a bad place have been omitted
to aid visibility.}
 \label{fig:same-dec-with-test}

\end{minipage}

\vspace{5mm}

\begin{minipage}{7cm}

We wish to model that players learn about previous
decisions of other players
by communication.
 To this end, we use the \emph{unfolding} of the net, 
where each place that is reachable via several transition paths is
duplicated into several copies of the place, each one representing its causal past.
The \emph{unfolding} of a game $\mathcal G$ 
is the unfolding of the underlying net $\mathcal N$,
denoted by  the branching process
$\beta_U = (\mathcal N^U,\fb)$,
where $\mathcal N^U$ is an occurrence net and 
$\fb$ is an initial homomorphism from $\mathcal N^U$
to $\mathcal N$, which ``labels'' the places and transitions
of $\mathcal N^U$ with the places and transitions of
$\mathcal N$.
In the graphic representation of games and unfoldings 
gray places denote elements of $\places_S$ and white places
elements of~$\places_E$.

\begin{example}
\label{exam:unfold-same-dec-with-test}
Fig.~\ref{fig:unfold-same-dec-with-test} shows the 
unfolding of the Petri game in Fig.~\ref{fig:same-dec-with-test}.
\qed
\end{example}

\end{minipage}
\hspace{7mm}
\begin{minipage}{8cm}


\begin{center}

 \psset{unit=1}

\scalebox{0.7}{
\begin{pspicture}(0,11.5)(10,19.5)  
   \psset{arrows=->,nodesep=.25cm}


\pscircle(5,19){.25} \pnode(5,19){01p0} 
    \rput(5,19.5){$\mathit{Env}$} 
    \rput(5,19){$\bullet$}

   
    \psframe(0.75,17.75)(1.25,18.25) \pnode(1,18){11t1} \rput(0.5,18){$t_1$}
    \psframe(8.75,17.75)(9.25,18.25) \pnode(9,18){12t2} \rput(9.5,18){$t_2$}


\pscircle(1,17){.25} \pnode(1,17){21p1} \rput(0.5,17){$A$ }

\pscircle[fillstyle=solid,fillcolor=lightgray](4,17.5){.25} 
   \pnode(4,17.5){02q0} \rput(4,18){$\mathit{Sys}$}
   \rput(4,17.5){$\bullet$}

\pscircle[fillstyle=solid,fillcolor=lightgray](6,17.5){.25} 
   \pnode(6,17.5){03q0} \rput(6,18){$\mathit{Sys}$} 
   \rput(6,17.5){$\bullet$}

\pscircle(9,17){.25} \pnode(9,17){22p2} \rput(9.6,17){$B$ }


\psframe(3.25,15.75)(3.75,16.25) 
   \pnode(3.5,16){t161} \rput(3.5,16.8){$t_1'$}

\psframe(4.25,15.75)(4.75,16.25) 
   \pnode(4.5,16){t162} \rput(4.6,16.6){$t_2'$}

\psframe(5.25,15.75)(5.75,16.25) 
   \pnode(5.5,16){t163} \rput(5.4,16.6){$t_1'$}

\psframe(6.25,15.75)(6.75,16.25) 
   \pnode(6.5,16){t164} \rput(6.5,16.8){$t_2'$}


\psframe[fillstyle=solid](0.75,15.25)(1.25,15.75) 
    \pnode(1,15.5){31tt1} \rput(0.3,15.5){$\mathit{test}_1$}
\psframe[fillstyle=solid](8.75,15.25)(9.25,15.75) 
    \pnode(9,15.5){32tt2} \rput(9.7,15.5){$\mathit{test}_2$}


\pscircle[fillstyle=solid,fillcolor=lightgray](3.5,15){.25} 
   \pnode(3.5,15){q151} \rput(3.2,15.4){$A'$ } 

\pscircle[fillstyle=solid,fillcolor=lightgray](4.5,15){.25} 
   \pnode(4.5,15){q152} \rput(4.2,15.4){$B'$ } 

\pscircle[fillstyle=solid,fillcolor=lightgray](5.5,15){.25} 
  \pnode(5.5,15){q153} \rput(5.9,15.4){$A'$ }

\pscircle[fillstyle=solid,fillcolor=lightgray](6.5,15){.25} 
  \pnode(6.5,15){q154} \rput(6.9,15.4){$B'$ }


\pscircle(1,14){.25} \pnode(1,14){51p1} 
    \rput(0.4,14){$EA$} 

\pscircle[fillstyle=solid,fillcolor=lightgray](2.1,14){.25} 
  \pnode(2.1,14){52q0} \rput(2.7,14){$\mathit{Sys}$} 

\pscircle[fillstyle=solid,fillcolor=lightgray](4.1,14){.25} 
  \pnode(4.1,14){53q0} \rput(4.1,14.5){$\mathit{Sys}$} 

\pscircle[fillstyle=solid,fillcolor=lightgray](5.85,14){.25} 
  \pnode(5.85,14){54q0} \rput(5.8,14.5){$\mathit{Sys}$} 

\pscircle[fillstyle=solid,fillcolor=lightgray](7.85,14){.25} 
  \pnode(7.85,14){55q0} \rput(7.2,14){$\mathit{Sys}$} 

\pscircle(9,14){.25} \pnode(9,14){56p2} 
    \rput(9.6,14){$EB$} 


\psframe(1.5,12.75)(2,13.25) 
   \pnode(1.75,13){61t11} \rput(1.65,13.6){$t_1'$}

\psframe(2.25,12.75)(2.75,13.25) 
   \pnode(2.5,13){61t21} \rput(2.6,13.6){$t_2'$}

\psframe(3.5,12.75)(4,13.25) 
   \pnode(3.75,13){62t11} \rput(3.65,13.6){$t_1'$}

\psframe(4.25,12.75)(4.75,13.25) 
   \pnode(4.5,13){62t21} \rput(4.6,13.6){$t_2'$}

\psframe(5.25,12.75)(5.75,13.25) 
   \pnode(5.5,13){63t11} \rput(5.4,13.6){$t_1'$}

\psframe(6,12.75)(6.5,13.25) 
   \pnode(6.25,13){63t21} \rput(6.35,13.6){$t_2'$}

\psframe(7.25,12.75)(7.75,13.25) 
   \pnode(7.5,13){64t11} \rput(7.4,13.6){$t_1'$}

\psframe(8,12.75)(8.5,13.25) 
   \pnode(8.25,13){64t21} \rput(8.35,13.6){$t_2'$}


\pscircle[fillstyle=solid,fillcolor=lightgray](1.75,12){.25} 
   \pnode(1.75,12){71q1} \rput(1.75,11.5){$A'$ } 

\pscircle[fillstyle=solid,fillcolor=lightgray](2.5,12){.25} 
   \pnode(2.5,12){71q2} \rput(2.5,11.5){$B'$ }

\pscircle[fillstyle=solid,fillcolor=lightgray](3.75,12){.25} 
   \pnode(3.75,12){72q1} \rput(3.75,11.5){$A'$ }

\pscircle[fillstyle=solid,fillcolor=lightgray](4.5,12){.25} 
   \pnode(4.5,12){72q2} \rput(4.5,11.5){$B'$ }

\pscircle[fillstyle=solid,fillcolor=lightgray](5.5,12){.25} 
   \pnode(5.5,12){73q1} \rput(5.5,11.5){$A'$ } 

\pscircle[fillstyle=solid,fillcolor=lightgray](6.25,12){.25} 
   \pnode(6.25,12){73q2} \rput(6.25,11.5){$B'$ }

\pscircle[fillstyle=solid,fillcolor=lightgray](7.5,12){.25} 
  \pnode(7.5,12){74q1} \rput(7.5,11.5){$A'$ }

\pscircle[fillstyle=solid,fillcolor=lightgray](8.25,12){.25} 
  \pnode(8.25,12){74q2} \rput(8.25,11.5){$B'$ }


 \ncline[offset=0mm]{01p0}{11t1}  
     \ncline[offset=0mm]{01p0}{12t2}
 \ncline[offset=0mm]{02q0}{31tt1}
     \ncline[offset=0mm]{02q0}{32tt2}
 \ncline[offset=0mm]{03q0}{31tt1}
     \ncline[offset=0mm]{03q0}{32tt2}

 \ncline[offset=0mm,linestyle=dashed,linecolor=blue]{02q0}{t161}
 \ncline[offset=0mm,linestyle=dashed,linecolor=blue]{02q0}{t162}

 \ncline[offset=0mm,linestyle=dashed,linecolor=blue]{03q0}{t163}
 \ncline[offset=0mm,linestyle=dashed,linecolor=blue]{03q0}{t164}

 \ncline[offset=0mm,linestyle=dashed,linecolor=blue]{t161}{q151}
 \ncline[offset=0mm,linestyle=dashed,linecolor=blue]{t162}{q152}
 \ncline[offset=0mm,linestyle=dashed,linecolor=blue]{t163}{q153}
 \ncline[offset=0mm,linestyle=dashed,linecolor=blue]{t164}{q154}

 \ncline[offset=0mm]{11t1}{21p1} 
     \ncline[offset=0mm]{12t2}{22p2}
 \ncline[offset=0mm]{21p1}{31tt1}
     \ncline[offset=0mm]{22p2}{32tt2}

 \ncline[offset=0mm]{31tt1}{51p1} 
 \ncline[offset=0mm]{31tt1}{52q0}
 \ncline[offset=0mm]{31tt1}{53q0}

 \ncline[offset=0mm]{32tt2}{54q0}
 \ncline[offset=0mm]{32tt2}{55q0}
 \ncline[offset=0mm]{32tt2}{56p2}

 \ncline[offset=0mm]{52q0}{61t11}
  \ncline[offset=0mm,linestyle=dashed,linecolor=blue]{52q0}{61t21}
 \ncline[offset=0mm]{53q0}{62t11}
   \ncline[offset=0mm,linestyle=dashed,linecolor=blue]{53q0}{62t21}

 \ncline[offset=0mm]{54q0}{63t21}
  \ncline[offset=0mm,linestyle=dashed,linecolor=blue]{54q0}{63t11}
 \ncline[offset=0mm]{55q0}{64t21}
   \ncline[offset=0mm,linestyle=dashed,linecolor=blue]{55q0}{64t11}

 \ncline[offset=0mm]{61t11}{71q1}
  \ncline[offset=0mm,linestyle=dashed,linecolor=blue]{61t21}{71q2}
 \ncline[offset=0mm]{62t11}{72q1}
   \ncline[offset=0mm,linestyle=dashed,linecolor=blue]{62t21}{72q2}

 \ncline[offset=0mm,linestyle=dashed,linecolor=blue]{63t11}{73q1}
  \ncline[offset=0mm]{63t21}{73q2}
 \ncline[offset=0mm,linestyle=dashed,linecolor=blue]{64t11}{74q1}
  \ncline[offset=0mm]{64t21}{74q2}

\end{pspicture}
} 

\end{center}


\caption{Unfolding of the Petri game 
in Fig.~\ref{fig:same-dec-with-test}. 
If the transitions shown with dashed lines are removed from the
unfolding, the resulting net represents a winning strategy for the system players, i.e., on the left-hand side, the system players
choose $A'$, and on the right-hand side, the system players
choose $B'$.}
 \label{fig:unfold-same-dec-with-test}

\end{minipage}

\end{figure}


A  global strategy is obtained from the unfolding by deleting 
some of the branches that are under control of the system players.
We call this a ``global'' strategy because it looks at all players simultaneously.
Note that nevertheless a strategy describes 
for each place which transitions
the player in that place can take.
Formally, this is expressed by the net-theoretic notion of subprocess.


An \emph{unfolded} (\emph{global}) 
\emph{strategy} for the system players 
in 
$\mathcal G$ 
is a subprocess $\sigma = (\mathcal{N}^\sigma, \fb^\sigma)$ of the unfolding $\beta_U = (\mathcal{N}^U, \fb)$ of $\mathcal N$
subject to the following conditions for all $p \in \places^\sigma$:

\begin{enumerate}

\item[(S1)] if $p \in \places_S^\sigma$ then
           $\sigma$ is deterministic at $p$,

\item[(S2)] if $p \in \places_E^\sigma$ then
         $\forall t \in \transitions^U: 
           (p,t) \in \flow^U \land$ 
           $ |pre^U(t)| = 1 \Rightarrow (p,t) \in \flow^\sigma$,
          i.e., at an environment place
          the strategy does not restrict any local transitions.

\end{enumerate}
Here $\places_S^\sigma = \places^\sigma \cap \fb^{-1}(\places_S)$
denotes the system places  
and
$\places_E^\sigma = \places^\sigma \cap \fb^{-1}(\places_E)$
the environment places in~$\places^\sigma$.
A strategy $\sigma$ is \emph{deterministic at a place} $p$ if
for all $M \in  \mathcal{R}( \mathcal{N}^\sigma)$, the
set of reachable markings in $\mathcal{N}^\sigma$:
\[
p \in M \Rightarrow
 \exists^{\le 1}\, t \in \transitions^\sigma:
 p \in pre(t) \subseteq M.
\]
Due to the unfolding,  
a decision taken by $\sigma$ in a place $p$ depends on the
causal past of $p$, which may be arbitrarily large.
The adjective ``global'' indicates that $\sigma$ looks
at all players simultaneously.
Local controllers are discussed in Section~\ref{section:distribution}.

\begin{example}
\label{exam:global-strategy}
Fig.~\ref{fig:unfold-same-dec-with-test} shows also
a global strategy for the system players of the Petri game in Fig.~\ref{fig:same-dec-with-test}.
\qed
\end{example}


A (\emph{concurrent}) \emph{play} of a Petri game  $\mathcal G$ 
is an initial concurrent run $\pi$ of the underlying net $\mathcal N$. 
If $\pi$ contains a place of $\bad$, the  \emph{environment wins}~$\pi$. 
Otherwise, the \emph{system players win} $\pi$.
Note that up to isomorphism we can assume that $\pi$ is a subprocess of
the unfolding $\beta^U$.
A play $\pi$ \emph{conforms to} a strategy~$\sigma$ 
if $\pi$ is a subprocess of $\sigma$.
A strategy $\sigma$ for the system players is \emph{winning} if 
the system players win every play that conforms to~$\sigma$.

Since the winning condition of a game is a {\emph{safety objective},
the system players can satisfy it by doing nothing.
To avoid such trivial solutions, we look for strategies $\sigma$
that are \emph{deadlock avoiding} in the sense that
$\forall\, M \in  \mathcal{R}( \mathcal{N}^\sigma):$
$ \exists\, t \in \transitions^U: pre(t) \subseteq M   
       \Rightarrow   
   \exists\, t \in \transitions^\sigma: pre(t) \subseteq M$,
i.e., if the unfolding can execute a transition 
the strategy $\sigma$ can as well, thus avoiding unnecessary deadlocks.
A marking where there is no enabled transition in the unfolding either is not a deadlock. Then we 
say that the game has \emph{terminated}.


A (\emph{global}) 
\emph{strategy} for the  system players in $\mathcal G$ 
is a pair $\sigma = (\mathcal{N}^\sigma, h^\sigma)$
consisting of a safe net $\mathcal{N}^\sigma$
and an initial homomorphism  $h^\sigma$ from
$\mathcal{N}^\sigma$ to $\mathcal N$
that is injective on transitions with the same preset, i.e.,
$\forall\, t_1, t_2 \in \transitions^U:
       \pre{t_1} = \pre{t_2} \land \fb(t_1) = \fb(t_2)$
       implies $t_1 = t_2$,
subject to the conditions (S1) and (S2) above.
A global strategy $\sigma$ may have cycles
and thus be finite, i.e., have a finite set 
$\places^\sigma \cup \transitions^\sigma$.

\section{Distribution}
\label{section:distribution}

We show that for Petri games with
a concurrency preserving underlying net, 
every global strategy $\sigma$ is distributable over local controllers.
A net $\mathcal N$ is \emph{concurrency preserving} if
every  transition $t \in \transitions$ satisfies $|pre(t)| = |post(t)|$.
The \emph{parallel composition} $\mathcal{N}_1\, ||\, \mathcal{N}_2$
of two nets
$\mathcal N_i = (\places_i, \transitions_i, \flow_i, {\initialmarking}_i)$, $i=1,2$, 
with $\places_1 \cap \places_2 = ß\emptyset$
is defined as the Petri net
$\mathcal{N}_1\, ||\, \mathcal{N}_2 =
(\places_1 \cup \places_2, 
\transitions_1 \cup \transitions_2, 
\flow_1 \cup \flow_2, 
{\initialmarking}_1 \cup {\initialmarking}_2)$
obtained by taking the componentwise union. The two nets 
synchronize on each common transition $t \in \transitions_1 \cap \transitions_2$
as in the process algebra CSP \cite{Hoa85,Old91}.

Let $\mathcal N = (\places,\transitions,\flow,\initialmarking)$
be a concurrency preserving, safe net
with the places partitioned into system and environment
places $\places=\places_S \cup \places_E$.
A \emph{slice} of $\mathcal N$ describes the course of
one token in  $\mathcal N$.
Formally, it is a net
$S = (\places^S,\transitions^S,\flow^S,\initialmarking^S)$,
where
$\places^S \subseteq \places_S$ or
     $\places^S \subseteq \places_E$,
$\transitions^S \subseteq \transitions$,
$\flow^S \subseteq \flow$, 
$\initialmarking^S \subseteq \initialmarking$
are minimal subsets satisfying

\begin{itemize}

\item $|\initialmarking^S| = 1$ 
         and 
         $\forall p \in \places^S: 
          post^N(p) \subseteq \transitions^S$
          and
          $\forall t \in \transitions^S:
          |pre^S(t)| = |post^S(t)| = 1$,

\item $\flow^S = 
      \flow \upharpoonright 
       (\places^S \times \transitions^S) \cup 
       (\transitions^S \times \places^S)$.

\end{itemize}
The net $\mathcal N$ is called \emph{reachable}
if every place and transition of $\mathcal N$
is reachable from its initial marking.


\begin{lemma}[Parallel Composition of Slices]
\label{lem:parcomp-slices}
Every safe reachable net $\mathcal N$
which is concurrency preserving 
is the parallel composition of slices:
$\mathcal{N} =  
  \|_{\, S \in \mathcal{S}}\ \mathcal{S},$
where $\mathcal S$ is a family of slices of $\mathcal N$ such that
$\{\places^S \mid S \in \mathcal{S} \}$ is a partition of $\places$.
\end{lemma}


A  \emph{local controller} specifies the moves of a single player
in a Petri game.
It is a pair 
$\mathcal C =(\mathcal N^C, h^C)$
consisting of a safe net $\mathcal N^C$ with one token,
i.e., $|\initialmarking^C| = 1$ and
$\forall t \in \transitions^C:  |pre^C(t)| = |post^C(t)| = 1$,
and a \emph{weak homomorphism} 
$h^C$ from $\mathcal{N}^C$ to $\mathcal N$, 
the underlying net of the Petri game.
A local controller $\mathcal C$ is \emph{finite} if
$\places^C \cup \transitions^C$ is a finite set.
It may have nondeterministic choices of transitions 
that are resolved (later) by synchronization with other
controllers working in parallel.
Unfolding $\mathcal N^C$ yields a branching process
$\beta^C = (\mathcal N^{CU}, \fb^C)$, where
$\fb^C$ is an initial homomorphism from $\mathcal N^{CU}$ to $\mathcal N^C$.
Then $\mathcal C^U = (\mathcal N^{CU}, h^C \circ \fb^C)$
is an \emph{unfolded local controller}.

A (n unfolded) strategy $\sigma$ is \emph{distributable} if
 $\sigma$ can be represented as the parallel composition of
(unfolded) local controllers for the environment and the system
players in the sense that the reachable part of the
parallel composition is isomorphic to~$\sigma$.
Using Lemma~\ref{lem:parcomp-slices} we show:


\begin{lemma}[Distribution]
\label{lemma:distribution}
Every unfolded global strategy 
for a concurrency-preserving Petri game
is distributable.
\end{lemma}

\begin{example}
\label{exam:}
The global strategy of Fig.~\ref{fig:unfold-same-dec-with-test} 
can be distributed into the local controllers of Fig.~\ref{fig:same-dec-loc-ctrl}.
\qed
\end{example}


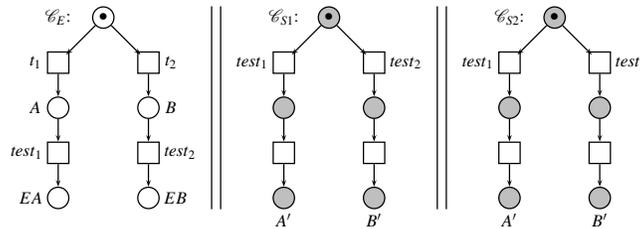
\begin{figure}[h]


\begin{center}

 \psset{unit=1}

\scalebox{0.6}{
\begin{pspicture}(0,3)(14,7)   
   \psset{nodesep=.25cm}


  \pnode(4.4,7.5){p447}
  \pnode(4.6,7.5){p467}

  \pnode(9.4,7.5){p947}
  \pnode(9.6,7.5){p967}

  
\pscircle[fillstyle=solid](2,7){.25} 
   \pnode(2,7){p27} 
   \rput(2,7){$\bullet$}
   \rput(1,7){$\mathcal{C}_E$:}

\pscircle[fillstyle=solid,fillcolor=lightgray](7,7){.25} 
   \pnode(7,7){p77} 
   \rput(7,7){$\bullet$}
   \rput(6,7){$\mathcal{C}_{S1}$:}

\pscircle[fillstyle=solid,fillcolor=lightgray](12,7){.25} 
   \pnode(12,7){p127} 
   \rput(12,7){$\bullet$}
   \rput(11,7){$\mathcal{C}_{S2}$:}


 \psframe(0.75,5.75)(1.25,6.25) \pnode(1,6){t16} 
     \rput(0.5,6){$t_1$}

 \psframe(2.75,5.75)(3.25,6.25) \pnode(3,6){t36} 
     \rput(3.5,6){$t_2$}

 \psframe(5.75,5.75)(6.25,6.25) \pnode(6,6){t66} 
     \rput(5.3,6){$\mathit{test}_1$}

 \psframe(7.75,5.75)(8.25,6.25) \pnode(8,6){t86} 
     \rput(8.7,6){$\mathit{test}_2$}

\psframe(10.75,5.75)(11.25,6.25) \pnode(11,6){t116} 
     \rput(10.3,6){$\mathit{test}_1$}

 \psframe(12.75,5.75)(13.25,6.25) \pnode(13,6){t136} 
     \rput(13.7,6){$\mathit{test}_2$}


\pscircle[fillstyle=solid](1,5){.25} 
   \pnode(1,5){p15}  \rput(0.5,5){$A$}

\pscircle[fillstyle=solid](3,5){.25} 
   \pnode(3,5){p35}  \rput(3.5,5){$B$}

\pscircle[fillstyle=solid,fillcolor=lightgray](6,5){.25} 
   \pnode(6,5){p65}  

 \pscircle[fillstyle=solid,fillcolor=lightgray](8,5){.25} 
   \pnode(8,5){p85} 

\pscircle[fillstyle=solid,fillcolor=lightgray](11,5){.25} 
   \pnode(11,5){p115}  

 \pscircle[fillstyle=solid,fillcolor=lightgray](13,5){.25} 
   \pnode(13,5){p135} 


 \psframe(0.75,3.75)(1.25,4.25) \pnode(1,4){t14} 
    \rput(0.3,4){$\mathit{test}_1$}

 \psframe(2.75,3.75)(3.25,4.25) \pnode(3,4){t34} 
    \rput(3.7,4){$\mathit{test}_2$}

 \psframe(5.75,3.75)(6.25,4.25) \pnode(6,4){t64} 

 \psframe(7.75,3.75)(8.25,4.25) \pnode(8,4){t84} 

\psframe(10.75,3.75)(11.25,4.25) \pnode(11,4){t114} 

 \psframe(12.75,3.75)(13.25,4.25) \pnode(13,4){t134} 


\pscircle[fillstyle=solid](1,3){.25} 
   \pnode(1,3){p13}  \rput(0.4,3){$EA$}

\pscircle[fillstyle=solid](3,3){.25} 
   \pnode(3,3){p33}  \rput(3.6,3){$EB$}

\pscircle[fillstyle=solid,fillcolor=lightgray](6,3){.25} 
   \pnode(6,3){p63}  \rput(6,2.5){$A'$}

 \pscircle[fillstyle=solid,fillcolor=lightgray](8,3){.25} 
   \pnode(8,3){p83}  \rput(8,2.5){$B'$}

\pscircle[fillstyle=solid,fillcolor=lightgray](11,3){.25} 
   \pnode(11,3){p113}  \rput(11,2.5){$A'$}

 \pscircle[fillstyle=solid,fillcolor=lightgray](13,3){.25} 
   \pnode(13,3){p133}  \rput(13,2.5){$B'$}


   \pnode(4.4,2.5){p442}
   \pnode(4.6,2.5){p462}

   \pnode(9.4,2.5){p942}
   \pnode(9.6,2.5){p962}


\ncline[offset=0mm]{p447}{p442} 
\ncline[offset=0mm]{p467}{p462}

\ncline[offset=0mm]{p947}{p942} 
\ncline[offset=0mm]{p967}{p962}


 \psset{arrows=->}

 \ncline[offset=0mm]{p27}{t16} \ncline[offset=0mm]{p27}{t36}
 \ncline[offset=0mm]{p77}{t66} \ncline[offset=0mm]{p77}{t86}
 \ncline[offset=0mm]{p127}{t116} \ncline[offset=0mm]{p127}{t136}

 \ncline[offset=0mm]{t16}{p15} 
 \ncline[offset=0mm]{t36}{p35}
  
          \ncline[offset=0mm]{t66}{p65} 
          \ncline[offset=0mm]{t86}{p85}

                  \ncline[offset=0mm]{t116}{p115} 
                  \ncline[offset=0mm]{t136}{p135}

 \ncline[offset=0mm]{p15}{t14} 
 \ncline[offset=0mm]{p35}{t34}

          \ncline[offset=0mm]{p65}{t64} 
          \ncline[offset=0mm]{p85}{t84}

                   \ncline[offset=0mm]{p115}{t114} 
                   \ncline[offset=0mm]{p135}{t134}

 \ncline[offset=0mm]{t14}{p13} 
 \ncline[offset=0mm]{t34}{p33}

          \ncline[offset=0mm]{t64}{p63} 
          \ncline[offset=0mm]{t84}{p83}

                   \ncline[offset=0mm]{t114}{p113} 
                   \ncline[offset=0mm]{t134}{p133}
 
\end{pspicture}
} 

\end{center}


\caption{The local controllers 
$\mathcal{C}_E$ for the environment and 
$\mathcal{C}_{S1}$, $\mathcal{C}_{S2}$ 
for the system players
work in parallel and synchronize on the 
transitions $\mathit{test}_1$ and $\mathit{test}_2$.
Applying the parallel composition $\|$ 
to the three controller nets
yields the winning strategy of Fig.~\ref{fig:unfold-same-dec-with-test}.}
 \label{fig:same-dec-loc-ctrl}

\end{figure}

%


\begin{theorem}
If the system players in a bounded and concurrency preserving
Petri game  have a winning
strategy, then they have a finite \emph{distributable} winning strategy.
\end{theorem}

\section{Cuts}
\label{section:finiteness}

In an unfolded strategy $\sigma$, a decision taken by $\sigma$ 
in a place $p$ depends on the causal past of $p$, which may be arbitrarily large. 
Similar to model checking approaches based on net unfoldings~\cite{Esp94}, we use \emph{cuts} (maximal subset of pairwise concurrent places) as small summaries of the causal past.
The standard notion of cuts 
is, however, problematic for games with multiple players, because it collects places without regard for the
(possibly different) knowledge of the individual players about the causal past.
To solve this problem, we introduce a new kind of cut, called \emph{mcut}, which guarantees that
the system players can be considered to be perfectly informed about the environment
decisions.

Throughout this section, we consider a Petri game $\mathcal G$
with underlying net $\mathcal N$, 
unfolding $\beta_U = (\mathcal N^U, \fb)$, and an unfolded
strategy $\sigma = (\mathcal N^\sigma, \fb^\sigma)$,
so $\mathcal N^\sigma  \sqsubseteq \mathcal N^U$ and
$\fb^\sigma = 
\fb \upharpoonright (\places^\sigma \cup \transitions^\sigma)$.
Since in $\mathcal N^\sigma$ the nondeterminism
of $\mathcal N^U$ has been restricted,
we distinguish for a node 
$x \in \places^\sigma \cup \transitions^\sigma$
the postconditions $post^\sigma(x)$ and $post^U(x)$ taken
in the nets $\mathcal N^\sigma$ and $\mathcal N^U$,
respectively. Note that $\mathit{post}^\sigma(x) \subseteq \mathit{post}^U(x)$.
For preconditions we have $\mathit{pre}^\sigma(x) = \mathit{pre}^U(x)$.
Thus, while the postconditions of nodes may be different
in $\mathcal N^\sigma$ and $\mathcal N^U$, 
their preconditions are identical.


\begin{figure}[h]

\begin{minipage}{6.5cm}

\textbf{Futures, mcuts and ecuts.} 

 For a cut $C$ of an occurrence net let 
 $C^+ = \{x \in \places \cup \transitions \mid$ $\exists s \in C:\ s \leq x \}$, 
where $\leq$ denotes the 
reflexive \emph{causal predecessor} relation
given by $\flow^*$.
For a subnet $\mathcal N' \sqsubseteq \mathcal N^U$
and a cut $C$ of $\mathcal N'$ we write
$\mathcal N'_{C^+} = (\mathcal N' \upharpoonright C^+)[C]$.

\smallskip

Note that 
$(\mathcal N^U_{C^+},\ \fb \upharpoonright C^+)$
is an initial branching process
of the net $\mathcal N[\ms{\fb}{C}]$, which is like 
$\mathcal N$ but starts at the initial marking $\ms{\fb}{C}$.
For cuts $C$ and $C'$ we write $C \le C'$ if 
$\forall\, x \in C\ \exists\, y \in C': x \le y$,
and $C < C'$ if $C \le C'$ and $C \neq C'$.

\smallskip

The \emph{future in $\mathcal N^\sigma$} 
of a node $x$ in $\mathcal N^\sigma$ 
is the set
$
 \mathit{fut}^\sigma(x) = 
 \{ y \in \places^\sigma \cup \transitions^\sigma  \mid  x \le y \}.
$
A $p$-\emph{cut} is a cut containing the place $p$.

\smallskip

For an environment place $p \in \places^\sigma$ 
we introduce now
$mcut(p)$ as the w.r.t.\ $\le$ minimal $p$-cut $C$ 
such that for all places $q \in C$,
either the system players have \emph{maximally progressed} at $q$,
in the sense that any further system transition would require an additional
environment transition starting from place $p$,
or the future starting at $q$ does not depend on  the environment.

\end{minipage}
\hspace{7mm}
\begin{minipage}{9cm}


\begin{center}

 \psset{unit=1}

\scalebox{0.68}{
\begin{pspicture}(-2,9)(10,18.5)   
   \psset{nodesep=.25cm}

\rput(-0.5,18.5){$\sigma:$}

\pscircle(5,18){.25} \pnode(5,18){01p0} 
    \rput(5,18.5){$p_0$} 
    \rput(5,18){$\bullet$}

\pscircle[fillstyle=solid,fillcolor=lightgray](3,17){.25} 
   \pnode(3,17){02q0} \rput(2.4,17){$q_0$}
   \rput(3,17){$\bullet$}

\pscircle[fillstyle=solid,fillcolor=lightgray](7,17){.25} 
   \pnode(7,17){03q0} \rput(7.6,17){$q_0$} 
   \rput(7,17){$\bullet$}

   %
     \rput(1,17){$p$}
     \rput(-1.1,17){$mcut(p)=$}
     \psframe[linestyle=dashed,fillstyle=none,fillcolor=white](-0.1,16.5)(8,17.5)
   %

   %
      \rput(1,14){$q$}
      \rput(-1.3,14){$ecut(p,tt_1)=$}
      \psframe[linestyle=solid,fillstyle=none,fillcolor=white](-0.1,13.5)(5,14.5)
   %

   %
      \rput(-0.7,12){$mcut(q)=$} 
       \pnode(0.2,14.75){m11}
       \pnode(1.4,14.75){m12}

       \pnode(0.2,12.25){m21}    \pnode(0.2,12.5){mm21}
       \pnode(1.4,12.25){m22}    \pnode(1.15,12.5){mm22}
       \pnode(5.25,12.5){m23}  \pnode(5,12.75){mm23}

       \pnode(0.2,11.25){m31}     \pnode(0,11.5){mm31}
       \pnode(1.4,11.5){m32}     \pnode(1.4,11.5){mm32}
       \pnode(5.25,11.5){m33}   \pnode(5,11.25){mm33}

       \ncline[linestyle=dashed,offset=0mm]{m11}{m31} 
       \ncline[linestyle=dashed,offset=0mm]{m12}{m22} 
       \ncline[linestyle=dashed,offset=0mm]{mm22}{m23} 
       \ncline[linestyle=dashed,offset=0mm]{mm31}{m33}  
       \ncline[linestyle=dashed,offset=0mm]{mm23}{mm33}

   %
   
    \psframe(0.75,17.75)(1.25,18.25) \pnode(1,18){11t1} \rput(0.5,18){$t_1$}
    \psframe(8.75,17.75)(9.25,18.25) \pnode(9,18){12t2} \rput(9.5,18){$t_2$}

 \pscircle(1,17){.25} \pnode(1,17){21p1} \rput(0.5,17){$p_1$ } 
 \pscircle(9,17){.25} \pnode(9,17){22p2} \rput(9.6,17){$p_2$ }

\psframe(0.75,15.25)(1.25,15.75) 
    \pnode(1,15.5){31tt1} \rput(0.4,15.5){$tt_1$}
\psframe(8.75,15.25)(9.25,15.75) 
    \pnode(9,15.5){32tt2} \rput(9.6,15.5){$tt_2$}


\pscircle(1,14){.25} \pnode(1,14){51p1} \rput(0.5,14){$p_1$} 

\pscircle[fillstyle=solid,fillcolor=lightgray](2.5,14){.25} 
  \pnode(2.5,14){52q0} \rput(2,14){$q_0$} 

\pscircle[fillstyle=solid,fillcolor=lightgray](4,14){.25} 
  \pnode(4,14){53q0} \rput(3.4,14){$q_0$} 

\pscircle[fillstyle=solid,fillcolor=lightgray](6,14){.25} 
  \pnode(6,14){54q0} \rput(6.6,14){$q_0$} 

\pscircle[fillstyle=solid,fillcolor=lightgray](7.5,14){.25} 
  \pnode(7.5,14){55q0} \rput(8,14){$q_0$} 

\pscircle(9,14){.25} \pnode(9,14){56p2} \rput(9.5,14){$p_2$}

 %
 %

\psframe(2.25,12.75)(2.75,13.25) 
   \pnode(2.5,13){61t11} \rput(2,13){$t_1'$}
\psframe(3.75,12.75)(4.25,13.25) 
   \pnode(4,13){62t11} \rput(3.5,13){$t_1'$}
   
\psframe(5.75,12.75)(6.25,13.25) 
   \pnode(6,13){63t21} \rput(6.5,13){$t_2'$}
\psframe(7.25,12.75)(7.75,13.25) 
   \pnode(7.5,13){64t21} \rput(8,13){$t_2'$}

\pscircle[fillstyle=solid,fillcolor=lightgray](2.5,12){.25} 
   \pnode(2.5,12){71q1} \rput(2,12.1){$q_1$ } 

\pscircle[fillstyle=solid,fillcolor=lightgray](4,12){.25} 
   \pnode(4,12){72q1} \rput(3.5,12.1){$q_1$ }

\pscircle[fillstyle=solid,fillcolor=lightgray](6,12){.25} 
   \pnode(6,12){73q2} \rput(6.6,12.1){$q_2$ } 

\pscircle[fillstyle=solid,fillcolor=lightgray](7.5,12){.25} 
  \pnode(7.5,12){74q2} \rput(8.1,12.1){$q_2$ }

\psframe(0.75,10.75)(1.25,11.25) 
    \pnode(1,11){81t111} \rput(0.3,11){$t_{111}$}
 \psframe(8.75,10.75)(9.25,11.25) 
    \pnode(9,11){82t222} \rput(9.7,11){$t_{222}$}

\pscircle(1,10){.25} \pnode(1,10){91p0} \rput(1,9.5){$p_0$ }
\pscircle[fillstyle=solid,fillcolor=lightgray](2,10){.25} 
    \pnode(2,10){92q0} \rput(2,9.5){$q_0$ } 
\pscircle[fillstyle=solid,fillcolor=lightgray](3,10){.25} 
    \pnode(3,10){93q0} \rput(3,9.5){$q_0$ }

\pscircle[fillstyle=solid,fillcolor=lightgray](7,10){.25}   
    \pnode(7,10){94q0} \rput(7,9.5){$q_0$ } 
\pscircle[fillstyle=solid,fillcolor=lightgray](8,10){.25} 
    \pnode(8,10){95q0} \rput(8,9.5){$q_0$ }
\pscircle(9,10){.25} \pnode(9,10){96p0} \rput(9,9.5){$p_0$ }


 \ncline[arrows=->,offset=0mm]{01p0}{11t1}  \ncline[arrows=->,offset=0mm]{01p0}{12t2}
 \ncline[arrows=->,offset=0mm]{02q0}{31tt1} \ncline[arrows=->,offset=0mm]{02q0}{32tt2}
 \ncline[arrows=->,offset=0mm]{03q0}{31tt1} \ncline[arrows=->,offset=0mm]{03q0}{32tt2}

 \ncline[arrows=->,offset=0mm]{11t1}{21p1} \ncline[arrows=->,offset=0mm]{12t2}{22p2}
 \ncline[arrows=->,offset=0mm]{21p1}{31tt1} \ncline[arrows=->,offset=0mm]{22p2}{32tt2}

 \ncline[arrows=->,offset=0mm]{31tt1}{51p1} 
 \ncline[arrows=->,offset=0mm]{31tt1}{52q0}
 \ncline[arrows=->,offset=0mm]{31tt1}{53q0}

 \ncline[arrows=->,offset=0mm]{32tt2}{54q0}
 \ncline[arrows=->,offset=0mm]{32tt2}{55q0}
 \ncline[arrows=->,offset=0mm]{32tt2}{56p2}

 \ncline[arrows=->,offset=0mm]{52q0}{61t11}
 \ncline[arrows=->,offset=0mm]{53q0}{62t11}

 \ncline[arrows=->,offset=0mm]{54q0}{63t21}
 \ncline[arrows=->,offset=0mm]{55q0}{64t21}

 \ncline[arrows=->,offset=0mm]{61t11}{71q1}
 \ncline[arrows=->,offset=0mm]{62t11}{72q1}

 \ncline[arrows=->,offset=0mm]{63t21}{73q2}
 \ncline[arrows=->,offset=0mm]{64t21}{74q2}

 \ncline[arrows=->,offset=0mm]{51p1}{81t111} 
 \ncline[arrows=->,offset=0mm]{71q1}{81t111} 
 \ncline[arrows=->,offset=0mm]{72q1}{81t111} 

 \ncline[arrows=->,offset=0mm]{73q2}{82t222}
 \ncline[arrows=->,offset=0mm]{74q2}{82t222}
 \ncline[arrows=->,offset=0mm]{56p2}{82t222}

\ncline[arrows=->,offset=0mm]{81t111}{91p0} 
\ncline[arrows=->,offset=0mm]{81t111}{92q0} 
\ncline[arrows=->,offset=0mm]{81t111}{93q0} 

\ncline[arrows=->,offset=0mm]{82t222}{94q0} 
\ncline[arrows=->,offset=0mm]{82t222}{95q0} 
\ncline[arrows=->,offset=0mm]{82t222}{96p0}

\end{pspicture}
} 

\end{center}


\caption{Shown is an initial part of an unfolding.
Consider the places $p$ and $q$ both labeled with~$p_1$.
Then $\mathit{mcut}(p)$ contains the upper places labeled $p_1, q_0, q_0$
and $\mathit{ecut}(p,tt_1)$ contains the places labeled $p_1, q_0, q_0$ in the middle,
whereas  $\mathit{mcut}(q)$ contains the places labeled $p_1, q_1, q_1$,
with the system players maximally progressed. 
Both mcuts have only places of type 1.}
 \label{fig:new-mcut-ecut}

\end{minipage}

\end{figure}
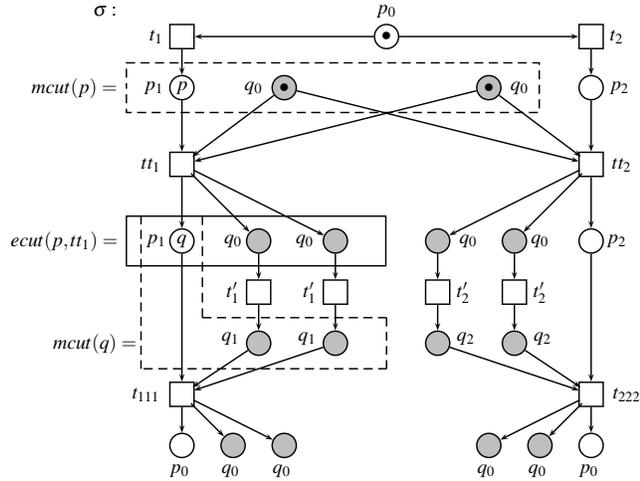


The formal definition is as follows: For a $p$-cut $C$ and a place $q \in C$ 
we define  $\mathit{type}(q)=1$  if
$
  \, \forall t \in post^\sigma(q): 
            (t \text{ reachable in } \mathcal N^\sigma_{C^+}  
            \Rightarrow p \le t)
$
and  $\mathit{type}(q)=2$   if
$
 \, \forall t \in \mathit{fut}^\sigma(q): 
            (t \text{ reachable in } \mathcal N^\sigma_{C^+}  
            \Rightarrow p \not \le t).
$
Note that $type(p) =1$.
By \emph{type-1(C)} we denote the set of all places in $C$ 
that have  type 1, and analogously for \emph{type-2(C)}.
Then we define:
$
 \mathit{mcut}(p) = \min_{\le} \{ C \mid 
          C \text{ is a } p\text{-cut of } \mathcal{N}^\sigma\  \land 
          \forall q \in C:  \mathit{type}(q) = 1 \lor  \mathit{type}(q) = 2 \}.
$
For an example, see Fig.~\ref{fig:new-mcut-ecut}.
 %
 %
 
 \begin{lemma}[Existence of mcuts]
 For every environment place $p \in \places^\sigma$,
 $\mathit{mcut}(p)$ is well-defined.
 \end{lemma}

 
 An \emph{ecut} results from an mcut by firing a single \emph{environment} transition.  
 Formally, given an environment place $p \in \places^\sigma$ 
 and a transition $t \in post^\sigma(p)$  with 
 environment participation 
 let $\mathit{ecut}(p,t)$ be the cut $C$ obtained by  firing $t$ at $\mathit{mcut}(p)$,
 formally $\mathit{mcut}(p) \fire{t}C$. 
 For an example, see Fig.~\ref{fig:new-mcut-ecut}.


\section{Deciding Petri Games}
\label{section:algorithm}

We now reduce Petri games to games over finite graphs, which can subsequently be solved by a standard fixed point construction.  Unlike the Petri game, the finite-graph game has only two players, Player~0 and Player 1, which both act on complete information. We construct a finite-graph game that is equivalent to the Petri game in the sense that the system players have a deadlock-avoiding and winning strategy in the Petri game iff Player~0 has a winning strategy in the finite-graph game. The key idea is that Player~1, representing the environment, is only allowed to make a decision at mcuts, which guarantees that the system players learn about the decision before they have to make their next choice. In this way, the system players can be considered to be perfectly informed.

A \emph{finite-graph game} $(V, V_0, V_1, I, E, W_0, W_1)$ consists of a finite set $V=V_0 \cup V_1$ of states, partitioned into Player~0's states $V_0$ and Player~1's states $V_1$, a set of initial states $I\subseteq V$, an edge relation $E \subseteq V \times V$, and disjoint sets of winning states $W_0, W_1 \subseteq V$ for Player~0 and Player~1, respectively.  A play is a possibly infinite sequence of states, constructed by letting Player~0 choose the next state from the $E$-successors whenever the play is in $V_0$ and letting Player~1 choose otherwise. Player~0 wins if the play reaches $W_0$ or forever avoids visiting $W_1$.

A \emph{strategy} for Player~0 is a function $f: V^*\cdot V_0
\rightarrow V$ that maps a prefix of a play ending in a state
owned by Player~0, i.e., a sequence of states that ends in a $V_0$
state, to some successor state according to $E$.  A
play \emph{conforms to} a strategy $f$, if all successors of $V_0$
states in the play are chosen according to $f$. A strategy is \emph{winning} for Player~0 if there is an initial state $v_0 \in I$ such that all plays that start in $v_0$ and conform to $f$ are won by Player~0.

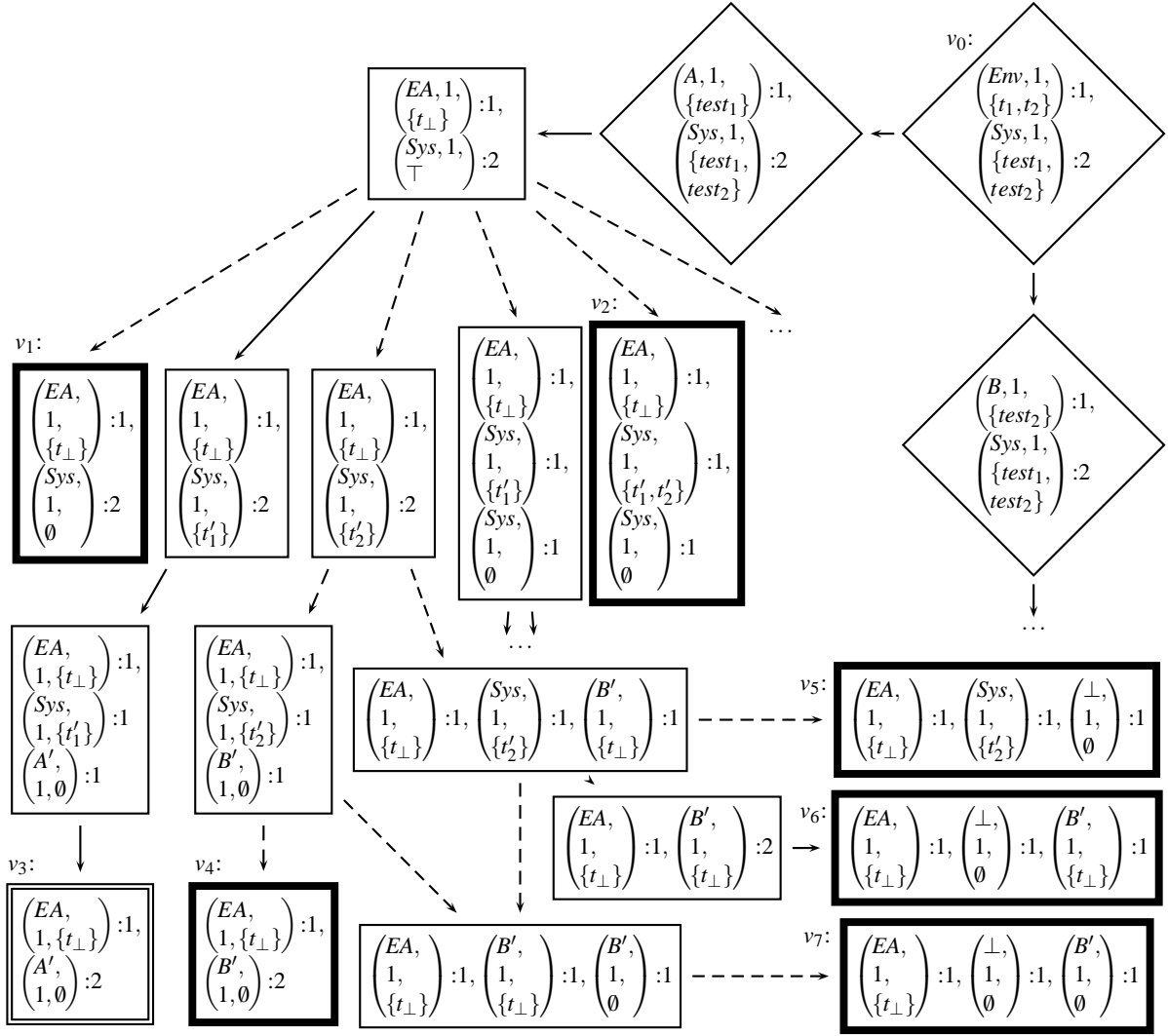
\begin{figure}[th]
\centering


 \psset{unit=1}

{\footnotesize
\begin{pspicture}(-0.7,-6.75)(14.625,7.5)  

   \psset{arrows=->,nodesep=.15cm}
\newcommand\nn{\hspace*{-0.2cm}}
\psdiamond(13,5.5)(1.8,1.8)\pnode(13,5.5){p31}\rput(13,5.5){$\begin{array}{l}\left(\nn\begin{array}{l}\mathit{Env}, 1,\\ \{t_1, t_2\}\end{array}\nn\right){:}1,\\ \left(\nn\begin{array}{l}\mathit{Sys}, 1,\\ \{test_1,\\ test_2\}\end{array}\nn\right){:}2\end{array}$}
\rput(12,6.75){$v_0$:}

\rput(5,5.5){\rnode{p21}{\psframebox{$\begin{array}{l}\left(\nn\begin{array}{l} \mathit{EA}, 1,\\ \{t_\bot\}\end{array}\nn\right){:}1,\\ \left(\nn\begin{array}{l}\mathit{Sys},1, \\ \top \end{array}\nn\right){:}2\end{array}$}}}

\psdiamond(8.875,5.5)(1.8,1.8)\pnode(8.875,5.5){p21x}\rput(8.875,5.5){$\begin{array}{l}\left(\nn\begin{array}{l} A, 1,\\ \{\mathit{test}_1\}\end{array}\nn\right){:}1,\\ \left(\nn\begin{array}{l}\mathit{Sys},1, \\ \{\mathit{test}_1,\\ \mathit{test}_2\}\end{array}\nn\right){:}2\end{array}$}


\psdiamond(13,1.25)(1.8,1.8)\pnode(13,1.25){p22}\rput(13,1.25){$\begin{array}{l}\left(\nn\begin{array}{l} B, 1,\\ \{\mathit{test}_2\}\end{array}\nn\right){:}1,\\ \left(\nn\begin{array}{l}\mathit{Sys},1,\\ \{\mathit{test}_1,\\ \mathit{test}_2\}\end{array}\nn\right){:}2\end{array}$}

\pnode(13,-1.25){q}
\rput(13,-1.25){$\ldots$}

\rput(-0.7,2.6){$v_1$:}
\rput(0,1){\rnode{p11}{\psframebox[linewidth=3pt]{$\nn\begin{array}{l}\left(\nn\begin{array}{l} EA,\\ 1,\\ \{t_\bot\} \end{array}\nn\right){:}1,\\ \left(\nn\begin{array}{l}\mathit{\mathit{Sys}},\\ 1,\\ \emptyset\end{array}\nn\right){:}2\end{array}\nn$}}}
\pnode(0,2.45){p11t}

\rput(2,1){\rnode{p12}{\psframebox{$\nn\begin{array}{l}\left(\nn\begin{array}{l} EA,\\ 1,\\ \{t_\bot\} \end{array}\nn\right){:}1,\\ \left(\nn\begin{array}{l}\mathit{Sys},\\ 1,\\ \{t_1'\}\end{array}\nn\right){:}2\end{array}\nn$}}}
\pnode(2,2.3){p12t}

\rput(4,1){\rnode{p13}{\psframebox{$\nn\begin{array}{l}\left(\nn\begin{array}{l} \mathit{EA},\\ 1,\\ \{t_\bot\} \end{array}\nn\right){:}1,\\ \left(\nn\begin{array}{l}\mathit{Sys},\\ 1,\\ \{t_2'\}\end{array}\nn\right){:}2\end{array}\nn$}}}
\pnode(4,2.3){p13t}

\rput(6,1){\rnode{p14}{\psframebox{$\nn\begin{array}{l}\left(\nn\begin{array}{l} \mathit{EA},\\ 1,\\ \{t_\bot\} \end{array}\nn\right){:}1,\\ \left(\nn\begin{array}{l}\mathit{Sys},\\ 1,\\ \{t_1'\}\end{array}\nn\right){:}1,\\ \left(\nn\begin{array}{l}\mathit{Sys},\\ 1,\\ \emptyset\end{array}\nn\right){:}1\end{array}\nn$}}}
\pnode(6,2.85){p14t}

\rput(8,1){\rnode{p15}{\psframebox[linewidth=3pt]{$\nn\begin{array}{l}\left(\nn\begin{array}{l} \mathit{EA},\\ 1,\\ \{t_\bot\} \end{array}\nn\right){:}1,\\ \left(\nn\begin{array}{l}\mathit{Sys},\\ 1,\\ \{t_1', t_2'\}\end{array}\nn\right){:}1,\\ \left(\nn\begin{array}{l}\mathit{Sys},\\ 1,\\ \emptyset\end{array}\nn\right){:}1\end{array}\nn$}}}
\pnode(8,2.9){p15t}
\rput(7.15,3.15){$v_2$:}

\pnode(9.7,2.95){p16t}
\rput(9.55,2.8){$\ldots$}

\rput(0,-2.5){\rnode{p02}{\psframebox{$\nn\begin{array}{l}\left(\nn\begin{array}{l} EA,\\ 1,\{t_\bot\} \end{array}\nn\right){:}1,\\ \left(\nn\begin{array}{l}\mathit{Sys},\\1,\{t_1'\}\end{array}\nn\right){:}1\\ \left(\nn\begin{array}{l}A',\\ 1, \emptyset\end{array}\nn\right){:}1\end{array}\nn$}}}

\rput(-0.8,-4.51){$v_3$:}
\rput(0,-5.7){\rnode{p02x}{\psframebox[doubleline=true]{$\nn\begin{array}{l}\left(\nn\begin{array}{l} EA,\\ 1,\{t_\bot\} \end{array}\nn\right){:}1,\\ \left(\nn\begin{array}{l}A',\\ 1, \emptyset\end{array}\nn\right){:}2\end{array}\nn$}}}

\rput(1.75,-4.51){$v_4$:}
\rput(2.5,-2.5){\rnode{p03}{\psframebox{$\nn\begin{array}{l}\left(\nn\begin{array}{l} \mathit{EA},\\ 1, \{t_\bot\} \end{array}\nn\right){:}1,\\ \left(\nn\begin{array}{l}\mathit{Sys},\\1,\{t_2'\}\end{array}\nn\right){:}1\\ \left(\nn\begin{array}{l}B',\\ 1, \emptyset\end{array}\nn\right){:}1\end{array}\nn$}}}
\rput(2.5,-5.7){\rnode{p03x}{\psframebox[linewidth=3pt]{$\nn\begin{array}{l}\left(\nn\begin{array}{l} EA,\\ 1, \{t_\bot\} \end{array}\nn\right){:}1,\\ \left(\nn\begin{array}{l}B',\\ 1, \emptyset\end{array}\nn\right){:}2\end{array}\nn$}}}

\rput(8,-4.25){\rnode{p03c}{\psframebox{$\nn\begin{array}{l}\left(\nn\begin{array}{l} \mathit{EA},\\ 1,\\ \{t_\bot\} \end{array}\nn\right){:}1, \left(\nn\begin{array}{l}B',\\ 1,\\ \{t_\bot\} \end{array}\nn\right){:}2\end{array}\nn$}}}
\pnode(5,-1.8){p03bt}
\rput(6,-2.5){\rnode{p03b}{\psframebox{$\nn\begin{array}{l}\left(\nn\begin{array}{l} \mathit{EA},\\ 1,\\ \{t_\bot\} \end{array}\nn\right){:}1, \left(\nn\begin{array}{l}\mathit{Sys},\\ 1,\\ \{t_2'\}\end{array}\nn\right){:}1, \left(\nn\begin{array}{l}B',\\ 1,\\ \{t_\bot\}\end{array}\nn\right){:}1\end{array}\nn$}}}
\rput(10,-2.05){$v_5$:}

\rput(6,-6){\rnode{p03d}{\psframebox{$\nn\begin{array}{l}\left(\nn\begin{array}{l} \mathit{EA},\\ 1,\\ \{t_\bot\} \end{array}\nn\right){:}1, \left(\nn\begin{array}{l}B',\\ 1,\\ \{t_\bot\}\end{array}\nn\right){:}1, \left(\nn\begin{array}{l}B',\\ 1,\\ \emptyset\end{array}\nn\right){:}1\end{array}\nn$}}}
\pnode(4,-4.3){p03dt}
\rput(12.5,-6){\rnode{p03e}{\psframebox[linewidth=3pt]{$\nn\begin{array}{l}\left(\nn\begin{array}{l} \mathit{EA},\\ 1,\\ \{t_\bot\} \end{array}\nn\right){:}1, \left(\nn\begin{array}{l}\bot,\\ 1,\\ \emptyset\end{array}\nn\right){:}1,\left(\nn\begin{array}{l}B',\\ 1,\\ \emptyset\end{array}\nn\right){:}1\end{array}\nn$}}}

\rput(12.5,-2.5){\rnode{p03f}{\psframebox[linewidth=3pt]{$\nn\begin{array}{l}\left(\nn\begin{array}{l} \mathit{EA},\\ 1,\\ \{t_\bot\} \end{array}\nn\right){:}1, \left(\nn\begin{array}{l}\mathit{Sys},\\ 1,\\ \{t_2'\}\end{array}\nn\right){:}1,\left(\nn\begin{array}{l}\bot,\\ 1,\\ \emptyset\end{array}\nn\right){:}1\end{array}\nn$}}}

\rput(12.5,-4.25){\rnode{p03z}{\psframebox[linewidth=3pt]{$\nn\begin{array}{l}\left(\nn\begin{array}{l} \mathit{EA},\\ 1,\\ \{t_\bot\} \end{array}\nn\right){:}1, \left(\nn\begin{array}{l}\bot,\\ 1,\\ \emptyset\end{array}\nn\right){:}1,\left(\nn\begin{array}{l}B',\\ 1,\\ \{t_\bot\}\end{array}\nn\right){:}1\end{array}\nn$}}}

\rput(10.05,-5.5){$v_7$:}
\rput(10,-3.8){$v_6$:}

\pnode(5.8,-1.5){p04a}\rput(6,-1.5){$\ldots$}
\pnode(6.2,-1.5){p04b}


  

\ncline[nodesep=1.9cm]{p31}{p21x}
\ncline[nodesepA=1.9cm]{p21x}{p21}

\ncline[nodesep=1.9cm]{p31}{p22}

\ncline[linestyle=dashed]{p21}{p11t}
\ncline{p21}{p12t}
\ncline[linestyle=dashed]{p21}{p13t}
\ncline[linestyle=dashed]{p21}{p14t}
\ncline[linestyle=dashed]{p21}{p15t}
\ncline[linestyle=dashed]{p21}{p16t}

\ncline{p12}{p02}\ncline{p02}{p02x}
\ncline[linestyle=dashed]{p13}{p03}
\ncline[linestyle=dashed]{p03}{p03x}
\ncline[linestyle=dashed]{p14}{p04b}
\ncline[linestyle=dashed]{p14}{p04a}
\ncline[linestyle=dashed]{p15}{p05a}
\ncline[linestyle=dashed]{p15}{p05b}
\ncline[linestyle=dashed]{p16}{p06a}
\ncline[linestyle=dashed]{p16}{p06b}

\ncline[linestyle=dashed]{p13}{p03bt}
\ncline[linestyle=dashed]{p03b}{p03c}
\ncline[linestyle=dashed]{p03d}{p03e}

\ncline[linestyle=dashed]{p03}{p03d}
\ncline[linestyle=dashed]{p03b}{p03f}

\ncline[linestyle=dashed]{p03b}{p03d}
\ncline[linestyle=dashed]{p03c}{p03z}

\ncline[nodesepA=1.9cm]{p22}{q}


\end{pspicture}
}

 
\caption{Part of the finite-graph game corresponding to the Petri game from Fig.~\ref{fig:same-dec-with-test}.}

\label{fig:algorithm-graph-game}

\end{figure}

To simulate a Petri game $\mathcal G = (\places_S, \places_E, 
\transitions, \flow, \initialmarking, \bad)$, we build a finite-graph game where the states are multisets consisting of triples $(p, \mathit{type}, T)$, where $p \in \places$ is a place, $\mathit{type}$ is a type, i.e., 1 or 2, and $T\in 2^\transitions \cup \{ \top\}$ is either a set of transitions representing the transitions chosen by a token in $p$ or a special symbol $\top$, indicating that a new choice needs to be made. We call these multisets \emph{decision sets}. For $k$-bounded Petri games, we limit the cardinality of the decision sets to $|\places|\cdot k$. 
A state belongs to Player~1 if the decision set corresponds to an mcut, and to Player~0 otherwise; i.e., the states of Player~1 consist of all decision sets where there is no $\top$ symbol and the outgoing transitions from type-1 places are either disabled or have an environment place in their precondition, the states of Player~0 consist of all other decision sets. The game starts with some initial marking, which fixes an arbitrary classification of types, all outgoing transitions for the environment places and an arbitrary selection of transitions for the system places. 
When there is a $\top$ symbol, Player~0 makes a choice for the transition set.
In other situations, the game continues by Player~0 choosing transitions from system places and Player~1 choosing transitions that involve an environment place. The choices of both players are restricted  to the transitions allowed in the decision set. There is an additional restriction based on the type of the places, which we will discuss below. Whenever a transition has fired, Player~0 chooses a new set of transitions for the newly reached system places. (In environment places, all outgoing transitions are always allowed.)

If no more transitions from type-1 places are enabled, the game ends. If this is due to termination, or if the decision set includes type-2 places, 
Player~0 wins. Player~1 wins if the game ends due to deadlock, if nondeterminism is encountered (i.e., two separate transitions, or two separate instances of the same transition, are enabled that share some system place in their precondition and have no environment places in their precondition), or if a bad place is visited.

\begin{example}
Figure~\ref{fig:algorithm-graph-game} shows (a part of) the finite-graph game corresponding to the Petri game from Fig.~\ref{fig:same-dec-with-test}. In Fig.~\ref{fig:same-dec-with-test}, the transitions leading to bad places have been omitted. For the purposes of this example, we assume that there is one additional transition $t_\bot$, which takes one token each from places $\mathit{EA}$ and $B'$ and puts one token on the bad place $\bot$ and one token back on $\mathit{EA}$. States of Player~0 are shown as rectangles, states of Player~1 as diamonds. Winning states for Player~0 are shown with double lines, winning states for Player~1 with bold lines. In addition to the initial state $v_0$ shown in Fig.~\ref{fig:algorithm-graph-game}, the game has further initial states that are omitted here. Player~0 has a winning strategy from $v_0$ (following the edges shown with solid lines). In state $v_3$, Player~0 wins, because the game terminates. In states $v_1$ and $v_4$, Player~1 wins, because a deadlock is reached. In state $v_2$, Player~1 wins, because nondeterminism is encountered. In states $v_5, v_6,$ and $v_7$, Player~1 wins, because the bad place $\bot$ is reached.
\end{example}

The finite-graph game as described so far does not yet ensure the correctness of the classification of the places into types 1 and 2. We need to make sure that the Petri game can indeed continue from type-2 places without dependencies on the environment and without visits to bad places. For this purpose, we  identify, in a preprocessing step, the largest subset  $\mathcal D$ of the set of decision sets that consists of only those decision sets that are either terminating or have at least one transition from type-2 places to another decision set in $\mathcal D$ that does not contain a bad place. We restrict the game to $\mathcal D$ and only allow (for both players) transitions that originate from type-1 places.

\begin{lemma}[Reduction to Finite-Graph Games]
\label{lem:reduction}
The system players have a deadlock-avoiding winning strategy in the Petri game iff 
Player~0 has a winning strategy in the finite-graph game.
\end{lemma}

To prove Lemma~\ref{lem:reduction}, we translate a winning strategy
for Player~0 in the finite-graph game into a deadlock-avoiding winning
strategy for the system players in the Petri game and vice versa.

Given a winning strategy $f$ of the finite-graph game, we inductively build a strategy $\sigma$ for the Petri game following the tree structure given by the possible choices of the environment token. In this way, we construct for each environment place in $\sigma$ a unique mcut, and for each subsequent ecut a causal net connecting the type-1 places of the ecut to the next mcut. The strategy is deadlock-avoiding because the decision sets of mcuts with deadlocks are winning for Player~1. The strategy is winning, because the plays that conform to $f$ avoid bad places. If the play in the finite-graph game is infinite, then the play in the Petri game is also infinite, traversing an infinite sequence of mcuts. If the play in the finite-graph game is finite, then this may be due to termination, in which case the play in the Petri game terminates as well; otherwise, the play must have reached a decision set with type-2 places, from which the play in the Petri game continues infinitely. 

Given a deadlock-avoiding winning strategy $\sigma$ of the Petri game and a prefix $w \in V^*\cdot V_0$ of a play of the finite-graph game, we compute the choice $f(w)$ of the strategy for the finite-graph game by simulating $w$ in $\sigma$: starting with the initial marking and firing the transitions of $w$ in $\sigma$, we arrive at a cut of $\sigma$ which is not an mcut; we choose an arbitrary enabled system transition and choose the decision set of the resulting cut as $f(w)$. For a cut $C$ in $\sigma$, the decision set is the multiset 
$
 \mathit{dec}[C]=\{ (\fb(p), \mathit{type}(p), \fb(post^\sigma(p))) \mid p \in C \}
$. 
The resulting strategy $f$ is winning from the decision set of the initial cut of $\sigma$.

The size of the finite-graph game is exponential in the size of the Petri game; the Petri game can therefore be solved in single-exponential time. A matching lower bound follows from the EXPTIME-hardness of combinatorial games~\cite{DBLP:journals/siamcomp/StockmeyerC79}. 

\begin{theorem}[Game Solving]
\label{thm:decidability}
For bounded Petri games with one environment player and
a bounded number of system players, the question
whether the system players have a winning strategy
is EXPTIME-complete. If a winning strategy for the system players exists,
it can be  constructed in exponential time.
\end{theorem}

Although the reachability problem is decidable also for unbounded Petri nets~\cite{Mayr/81/algorithm}, we cannot decide unbounded Petri games. This is an immediate consequence
of the undecidability of VASS (Vector Addition Systems with States) games~\cite{Abdulla+others/03/Deciding}. 

\smallskip

\begin{theorem}
For unbounded Petri games, the question whether the system players have a winning strategy is undecidable.
\end{theorem}

\section{Related Work}
\label{section:related}

There is a significant body of work on synthesis and control based on
Petri nets (cf.~\cite{Giua/92/Petri,Zhou+others/95/Generation,Raskin03petrigames,Buy/05/Supervisory}).
These approaches differ from ours in that they solve supervisory
control problems or two-player games on the state space created by the
Petri net. Hence, these approaches solve the single-process synthesis
problem, as opposed to the multi-process synthesis problem for
concurrent systems considered in this paper.

For distributed systems, much work has focused on finding architectures for which the realizability question is decidable. Most research on this problem is in the setting of synchronous processes with \emph{shared-variable} communication, introduced by Pnueli and Rosner.  A general game model for these types of realizability problems are Walukiewicz and Mohalik's \emph{distributed games}~\cite{Walukiewicz+Mohalik/03/Distributed}. While undecidable in general~\cite{Pnueli+Rosner/90/Distributed}, the distributed synthesis problem can be solved  in the Pnueli/Rosner setting for a number of interesting architectures, including pipelines~\cite{Rosner/92/Modular}, rings~\cite{Kupferman+Vardi/01/Synthesizing}, and generally all architectures where the processes can be \emph{ordered} according to their informedness~\cite{Finkbeiner+Schewe/05/Distributed}. Unfortunately, all these decision procedures have nonelementary complexity.
For the asynchronous games based on Zielonka's automata, decidability has been also been established for specific classes of architectures such as trees~\cite{Genest+others/ICALP2013/Asynchronous}. 
Another important line of work concerns the alternating-time temporal
logics, which are interpreted over concurrent game
structures~\cite{Alur+Henziger+Kupferman/02/ATL}. The difference
between Petri games and these approaches is that Petri games link
informedness to causality 
instead of referring to a separate, static, specification of the relative
informedness in an architecture.

In the literature on Petri nets, 
unfoldings have been used 
conceptually to connect Petri net theory with event structures
\cite{NPW81,BF88,Eng91,MMS96}
and 
practically to obtain algorithms for deciding reachability.
These algorithms are based on constructing a finite canonical prefix of the
in general infinite net unfolding that contain all reachable markings
\cite{Esp94,KKV03,EH08}.
We use net unfoldings as a \emph{uniform conceptual basis} 
to define strategies and plays  
as well as suitable cuts for analyzing the strategies.
Net unfoldings enable us to formalize the intended degree of
informedness of each player at a given place: it is the causal past of that place,
concurrent activities beyond that past are not visible.
Such a  causal view is also chosen in \cite{Gastin+others/04/Distributed},
for the setting of Zielonka's automata \cite{Zielonka/95/Asynchronous}.

\section{Conclusions}
\label{section:conclusion}

We have introduced Petri games, an extension of Petri nets where the
tokens represent players who make individual, independent decisions.  
Using tokens as the carriers of information, Petri games
link information flow to causality: decisions may only use information
resulting from decisions that they also depend on causally. This makes
Petri games a convenient formalism to reason about asynchronous concurrent programs
as well as manufacturing cells~\cite{Zhou+others/95/Generation},
business work flows~\cite{Aal98}, and other distributed applications.
Our synthesis algorithm is applicable to Petri games where the number
of system tokens is bounded by some arbitrary number, and the number
of environment tokens is bounded by~1.  This leaves two important open
problems. 
The first open problem is whether Petri games with more than one environment token are
decidable; if so, what is the precise complexity? The decidability
result for tree architectures~\cite{Genest+others/ICALP2013/Asynchronous} is both encouraging and discouraging;
encouraging, because at least some architectures that are undecidable
in the Pnueli/Rosner setting are decidable for distributed systems
with causal memory. Discouraging, because the complexity of the
synthesis algorithm is nonelementary.
The second open problem is to find synthesis methods for unbounded Petri games. While we have shown that 
the problem is in general undecidable, it is an interesting challenge for future research to develop semi-algorithms for unbounded Petri games and to find other restrictions besides boundedness that make the synthesis problem decidable.

\bibliographystyle{eptcsini}
\bibliography{IEEEabrv,bib}

\begin{thebibliography}{10}
\providecommand{\bibitemdeclare}[2]{}
\providecommand{\surnamestart}{}
\providecommand{\surnameend}{}
\providecommand{\urlprefix}{Available at }
\providecommand{\url}[1]{\texttt{#1}}
\providecommand{\href}[2]{\texttt{#2}}
\providecommand{\urlalt}[2]{\href{#1}{#2}}
\providecommand{\doi}[1]{doi:\urlalt{http://dx.doi.org/#1}{#1}}
\providecommand{\bibinfo}[2]{#2}

\bibitemdeclare{article}{Aal98}
\bibitem{Aal98}
\bibinfo{author}{W.M.P.v. \surnamestart Aalst\surnameend}
  (\bibinfo{year}{1998}): \emph{\bibinfo{title}{The application of {Petri} nets
  to workflow management}}.
\newblock {\sl \bibinfo{journal}{J. of Circuits, Systems and Computers}}
  \bibinfo{volume}{8}, pp. \bibinfo{pages}{21--66},
  \doi{10.1142/S0218126698000043}.

\bibitemdeclare{inproceedings}{Abdulla+others/03/Deciding}
\bibitem{Abdulla+others/03/Deciding}
\bibinfo{author}{P.A. \surnamestart Abdulla\surnameend},
  \bibinfo{author}{A.~\surnamestart Bouajjani\surnameend} \&
  \bibinfo{author}{J.~\surnamestart d'Orso\surnameend} (\bibinfo{year}{2003}):
  \emph{\bibinfo{title}{Deciding Monotonic Games}}.
\newblock In: {\sl \bibinfo{booktitle}{Proc. CSL}}, {\sl
  \bibinfo{series}{LNCS}} \bibinfo{volume}{2803},
  \bibinfo{publisher}{Springer-Verlag}, pp. \bibinfo{pages}{1--14},
  \doi{10.1007/978-3-540-45220-1\_1}.

\bibitemdeclare{article}{Alur+Henziger+Kupferman/02/ATL}
\bibitem{Alur+Henziger+Kupferman/02/ATL}
\bibinfo{author}{R.~\surnamestart Alur\surnameend}, \bibinfo{author}{T.A.
  \surnamestart Henzinger\surnameend} \& \bibinfo{author}{O.~\surnamestart
  Kupferman\surnameend} (\bibinfo{year}{2002}):
  \emph{\bibinfo{title}{Alternating-time temporal logic}}.
\newblock {\sl \bibinfo{journal}{Journal of the ACM}}
  \bibinfo{volume}{49}(\bibinfo{number}{5}), pp. \bibinfo{pages}{672--713},
  \doi{10.1145/585265.585270}.

\bibitemdeclare{book}{BF88}
\bibitem{BF88}
\bibinfo{author}{E.~\surnamestart Best\surnameend} \&
  \bibinfo{author}{C.~\surnamestart Fern\'{a}ndez\surnameend}
  (\bibinfo{year}{1988}): \emph{\bibinfo{title}{Nonsequential Processes}}.
\newblock \bibinfo{publisher}{Springer}, \doi{10.1007/978-3-642-73483-0}.

\bibitemdeclare{article}{Buy/05/Supervisory}
\bibitem{Buy/05/Supervisory}
\bibinfo{author}{U.~\surnamestart Buy\surnameend},
  \bibinfo{author}{H.~\surnamestart Darabi\surnameend},
  \bibinfo{author}{M.~\surnamestart Lehene\surnameend} \&
  \bibinfo{author}{V.~\surnamestart Venepally\surnameend}
  (\bibinfo{year}{2005}): \emph{\bibinfo{title}{Supervisory Control of Time
  {Petri} Nets Using Net Unfolding}}.
\newblock {\sl \bibinfo{journal}{Annual International Computer Software and
  Applications Conference}} \bibinfo{volume}{2}, pp. \bibinfo{pages}{97--100},
  \doi{10.1109/COMPSAC.2005.148}.

\bibitemdeclare{article}{Eng91}
\bibitem{Eng91}
\bibinfo{author}{J.~\surnamestart Engelfriet\surnameend}
  (\bibinfo{year}{1991}): \emph{\bibinfo{title}{Branching processes of {Petri}
  nets}}.
\newblock {\sl \bibinfo{journal}{Acta Informatica}}
  \bibinfo{volume}{28}(\bibinfo{number}{6}), pp. \bibinfo{pages}{575--591},
  \doi{10.1007/BF01463946}.

\bibitemdeclare{article}{Esp94}
\bibitem{Esp94}
\bibinfo{author}{J.~\surnamestart Esparza\surnameend} (\bibinfo{year}{1994}):
  \emph{\bibinfo{title}{Model checking using net unfoldings}}.
\newblock {\sl \bibinfo{journal}{Science of Computer Programming}}
  \bibinfo{volume}{23}, pp. \bibinfo{pages}{151--195},
  \doi{10.1016/0167-6423(94)00019-0}.

\bibitemdeclare{book}{EH08}
\bibitem{EH08}
\bibinfo{author}{J.~\surnamestart Esparza\surnameend} \&
  \bibinfo{author}{K.~\surnamestart Heljanko\surnameend}
  (\bibinfo{year}{2008}): \emph{\bibinfo{title}{Unfoldings -- A Partial-Order
  Approach to Model Checking}}.
\newblock \bibinfo{publisher}{Springer}, \doi{10.1007/978-3-540-77426-6}.

\bibitemdeclare{inproceedings}{Finkbeiner+Schewe/05/Distributed}
\bibitem{Finkbeiner+Schewe/05/Distributed}
\bibinfo{author}{B.~\surnamestart Finkbeiner\surnameend} \&
  \bibinfo{author}{S.~\surnamestart Schewe\surnameend} (\bibinfo{year}{2005}):
  \emph{\bibinfo{title}{Uniform Distributed Synthesis}}.
\newblock In: {\sl \bibinfo{booktitle}{Proc. LICS}}, \bibinfo{publisher}{IEEE
  Computer Society Press}, pp. \bibinfo{pages}{321--330},
  \doi{10.1109/LICS.2005.53}.

\bibitemdeclare{inproceedings}{Gastin+others/04/Distributed}
\bibitem{Gastin+others/04/Distributed}
\bibinfo{author}{P.~\surnamestart Gastin\surnameend},
  \bibinfo{author}{B.~\surnamestart Lerman\surnameend} \&
  \bibinfo{author}{M.~\surnamestart Zeitoun\surnameend} (\bibinfo{year}{2004}):
  \emph{\bibinfo{title}{Distributed Games with Causal Memory Are Decidable for
  Series-Parallel Systems}}.
\newblock In: {\sl \bibinfo{booktitle}{Proc. FSTTCS}}, pp.
  \bibinfo{pages}{275--286}, \doi{10.1007/978-3-540-30538-5\_23}.

\bibitemdeclare{inproceedings}{Genest+others/ICALP2013/Asynchronous}
\bibitem{Genest+others/ICALP2013/Asynchronous}
\bibinfo{author}{B.~\surnamestart Genest\surnameend},
  \bibinfo{author}{H.~\surnamestart Gimbert\surnameend},
  \bibinfo{author}{A.~\surnamestart Muscholl\surnameend} \&
  \bibinfo{author}{I.~\surnamestart Walukiewicz\surnameend}
  (\bibinfo{year}{2013}): \emph{\bibinfo{title}{Asynchronous Games over Tree
  Architectures}}.
\newblock In: {\sl \bibinfo{booktitle}{Proc. ICALP'13, Part II}}, {\sl
  \bibinfo{series}{LNCS}} \bibinfo{volume}{7966},
  \bibinfo{publisher}{Springer}, pp. \bibinfo{pages}{275--286},
  \doi{10.1007/978-3-642-39212-2\_26}.

\bibitemdeclare{phdthesis}{Giua/92/Petri}
\bibitem{Giua/92/Petri}
\bibinfo{author}{A.~\surnamestart Giua\surnameend} (\bibinfo{year}{1992}):
  \emph{\bibinfo{title}{Petri Nets as Discrete Event Models for Supervisory
  Control}}.
\newblock Ph.D. thesis, \bibinfo{school}{Rensselaer Polytechnic Institute}.

\bibitemdeclare{book}{Hoa85}
\bibitem{Hoa85}
\bibinfo{author}{C.A.R. \surnamestart Hoare\surnameend} (\bibinfo{year}{1985}):
  \emph{\bibinfo{title}{Communicating Sequential Processes}}.
\newblock \bibinfo{publisher}{Prentice Hall}, \doi{10.1145/359576.359585}.

\bibitemdeclare{article}{KKV03}
\bibitem{KKV03}
\bibinfo{author}{V.~\surnamestart Khomenko\surnameend},
  \bibinfo{author}{M.~\surnamestart Koutny\surnameend} \&
  \bibinfo{author}{W.~\surnamestart Vogler\surnameend} (\bibinfo{year}{2003}):
  \emph{\bibinfo{title}{Canonical prefixes of {Petri} net unfoldings}}.
\newblock {\sl \bibinfo{journal}{Acta Informatica}} \bibinfo{volume}{40}, pp.
  \bibinfo{pages}{95--118}, \doi{10.1007/3-540-45657-0\_49}.

\bibitemdeclare{inproceedings}{Kupferman+Vardi/01/Synthesizing}
\bibitem{Kupferman+Vardi/01/Synthesizing}
\bibinfo{author}{O.~\surnamestart Kupferman\surnameend} \&
  \bibinfo{author}{M.Y. \surnamestart Vardi\surnameend} (\bibinfo{year}{2001}):
  \emph{\bibinfo{title}{Synthesizing Distributed Systems}}.
\newblock In: {\sl \bibinfo{booktitle}{Proc. LICS}}, \bibinfo{publisher}{IEEE
  Computer Society Press}, pp. \bibinfo{pages}{389--398},
  \doi{10.1109/LICS.2001.932514}.

\bibitemdeclare{inproceedings}{MadhusudanTY/05/MSO}
\bibitem{MadhusudanTY/05/MSO}
\bibinfo{author}{P.~\surnamestart Madhusudan\surnameend}, \bibinfo{author}{P.S.
  \surnamestart Thiagarajan\surnameend} \& \bibinfo{author}{S.~\surnamestart
  Yang\surnameend} (\bibinfo{year}{2005}): \emph{\bibinfo{title}{The {MSO}
  Theory of Connectedly Communicating Processes}}.
\newblock In: {\sl \bibinfo{booktitle}{Proc. FSTTCS'05}}, {\sl
  \bibinfo{series}{LNCS}} \bibinfo{volume}{3821},
  \bibinfo{publisher}{Springer}, pp. \bibinfo{pages}{201--212},
  \doi{10.1007/11590156\_16}.

\bibitemdeclare{inproceedings}{Mayr/81/algorithm}
\bibitem{Mayr/81/algorithm}
\bibinfo{author}{E.W. \surnamestart Mayr\surnameend} (\bibinfo{year}{1981}):
  \emph{\bibinfo{title}{An algorithm for the general Petri net reachability
  problem}}.
\newblock In: {\sl \bibinfo{booktitle}{Proc. 13th ACM STOC}},
  \bibinfo{publisher}{ACM}, pp. \bibinfo{pages}{238--246},
  \doi{10.1145/800076.802477}.

\bibitemdeclare{article}{MMS96}
\bibitem{MMS96}
\bibinfo{author}{J.~\surnamestart Meseguer\surnameend},
  \bibinfo{author}{U.~\surnamestart Montanari\surnameend} \&
  \bibinfo{author}{V.~\surnamestart Sassone\surnameend} (\bibinfo{year}{1996}):
  \emph{\bibinfo{title}{Process versus unfolding semantics for Place/Transition
  {Petri} nets}}.
\newblock {\sl \bibinfo{journal}{TCS}} \bibinfo{volume}{153}, pp.
  \bibinfo{pages}{171--210}, \doi{10.1016/0304-3975(95)00121-2}.

\bibitemdeclare{article}{NPW81}
\bibitem{NPW81}
\bibinfo{author}{M.~\surnamestart Nielsen\surnameend}, \bibinfo{author}{G.D.
  \surnamestart Plotkin\surnameend} \& \bibinfo{author}{G.~\surnamestart
  Winskel\surnameend} (\bibinfo{year}{1981}): \emph{\bibinfo{title}{Petri Nets,
  Event Structures and Domains, Part I}}.
\newblock {\sl \bibinfo{journal}{Theor. Comput. Sci.}} \bibinfo{volume}{13},
  pp. \bibinfo{pages}{85--108}, \doi{10.1016/0304-3975(81)90112-2}.

\bibitemdeclare{book}{Old91}
\bibitem{Old91}
\bibinfo{author}{E.R. \surnamestart Olderog\surnameend} (\bibinfo{year}{1991}):
  \emph{\bibinfo{title}{Nets, Terms and Formulas: Three Views of Concurrent
  Processes and Their Relationship}}.
\newblock \bibinfo{publisher}{Cambridge University Press},
  \doi{10.1017/CBO9780511526589}.

\bibitemdeclare{inproceedings}{Pnueli+Rosner/90/Distributed}
\bibitem{Pnueli+Rosner/90/Distributed}
\bibinfo{author}{A.~\surnamestart Pnueli\surnameend} \&
  \bibinfo{author}{R.~\surnamestart Rosner\surnameend} (\bibinfo{year}{1990}):
  \emph{\bibinfo{title}{Distributed Reactive Systems are Hard to Synthesize}}.
\newblock In: {\sl \bibinfo{booktitle}{Proc. {FOCS}}}, \bibinfo{publisher}{IEEE
  Computer Society Press}, pp. \bibinfo{pages}{746--757},
  \doi{10.1109/FSCS.1990.89597}.

\bibitemdeclare{techreport}{Raskin03petrigames}
\bibitem{Raskin03petrigames}
\bibinfo{author}{J.F. \surnamestart Raskin\surnameend},
  \bibinfo{author}{M.~\surnamestart Samuelides\surnameend} \&
  \bibinfo{author}{L.V. \surnamestart Begin\surnameend} (\bibinfo{year}{2003}):
  \emph{\bibinfo{title}{Petri Games are Monotone but Difficult to Decide}}.
\newblock \bibinfo{type}{Technical Report},
  \bibinfo{institution}{Universit{\'e} Libre De Bruxelles}.

\bibitemdeclare{book}{Rei85}
\bibitem{Rei85}
\bibinfo{author}{W.~\surnamestart Reisig\surnameend} (\bibinfo{year}{1985}):
  \emph{\bibinfo{title}{Petri Nets -- An Introduction}}.
\newblock \bibinfo{publisher}{Springer}, \doi{10.1007/978-3-642-69968-9}.

\bibitemdeclare{phdthesis}{Rosner/92/Modular}
\bibitem{Rosner/92/Modular}
\bibinfo{author}{R.~\surnamestart Rosner\surnameend} (\bibinfo{year}{1992}):
  \emph{\bibinfo{title}{Modular Synthesis of Reactive Systems}}.
\newblock Ph.D. thesis, \bibinfo{school}{Weizmann Institute of Science,
  Rehovot, Israel}.

\bibitemdeclare{article}{DBLP:journals/siamcomp/StockmeyerC79}
\bibitem{DBLP:journals/siamcomp/StockmeyerC79}
\bibinfo{author}{L.J. \surnamestart Stockmeyer\surnameend} \&
  \bibinfo{author}{A.K. \surnamestart Chandra\surnameend}
  (\bibinfo{year}{1979}): \emph{\bibinfo{title}{Provably Difficult
  Combinatorial Games}}.
\newblock {\sl \bibinfo{journal}{SIAM J. Comput.}}
  \bibinfo{volume}{8}(\bibinfo{number}{2}), pp. \bibinfo{pages}{151--174},
  \doi{10.1137/0208013}.

\bibitemdeclare{inproceedings}{Walukiewicz+Mohalik/03/Distributed}
\bibitem{Walukiewicz+Mohalik/03/Distributed}
\bibinfo{author}{I.~\surnamestart Walukiewicz\surnameend} \&
  \bibinfo{author}{S.~\surnamestart Mohalik\surnameend} (\bibinfo{year}{2003}):
  \emph{\bibinfo{title}{Distributed Games}}.
\newblock In: {\sl \bibinfo{booktitle}{Proc. FSTTCS'03}}, {\sl
  \bibinfo{series}{LNCS}} \bibinfo{volume}{2914}, pp.
  \bibinfo{pages}{338--351}, \doi{10.1007/978-3-540-24597-1\_29}.

\bibitemdeclare{article}{Zhou+others/95/Generation}
\bibitem{Zhou+others/95/Generation}
\bibinfo{author}{Q.~\surnamestart Zhou\surnameend},
  \bibinfo{author}{M.~\surnamestart Wang\surnameend} \& \bibinfo{author}{S.P.
  \surnamestart Dutta\surnameend} (\bibinfo{year}{1995}):
  \emph{\bibinfo{title}{Generation of optimal control policy for flexible
  manufacturing cells: A {Petri} net approach}}.
\newblock {\sl \bibinfo{journal}{Intern. Journal of Advanced Manufacturing
  Technology}} \bibinfo{volume}{10}, pp. \bibinfo{pages}{59--65},
  \doi{10.1007/BF01184279}.

\bibitemdeclare{incollection}{Zielonka/95/Asynchronous}
\bibitem{Zielonka/95/Asynchronous}
\bibinfo{author}{W.~\surnamestart Zielonka\surnameend} (\bibinfo{year}{1995}):
  \emph{\bibinfo{title}{Asynchronous Automata}}.
\newblock In \bibinfo{editor}{G.~\surnamestart Rozenberg\surnameend} \&
  \bibinfo{editor}{V.~\surnamestart Diekert\surnameend}, editors: {\sl
  \bibinfo{booktitle}{Book of Traces}}, \bibinfo{publisher}{World Scientific},
  pp. \bibinfo{pages}{205--248}, \doi{10.1142/9789814261456\_0007}.

\end{thebibliography}

\newpage
 

\end{document}